\documentclass[11pt]{article}
\pdfoutput=1 
\usepackage{shorthand}

\usepackage{mathtools}
\usepackage{booktabs}
\usepackage[english]{babel}
\usepackage{amsmath,amssymb,amsbsy,amstext, amsthm, simplewick, amsfonts}
\usepackage{graphicx}
\usepackage[small]{caption}
\usepackage{siunitx}
\usepackage{upgreek}
\usepackage{framed}
\usepackage{wrapfig}
\usepackage{multirow}
\usepackage{bbm}
\usepackage[numbers,sort&compress]{natbib}
\usepackage[svgnames,dvipsnames,x11names]{xcolor}
\usepackage[utf8x]{inputenc}
\usepackage{selinput}
\usepackage{bm}
\usepackage{float}
\usepackage{geometry}
\usepackage{yfonts}
\usepackage{caption}
\usepackage{subcaption}
\usepackage{sidecap}
\usepackage{longtable}
\usepackage{anyfontsize}

\usepackage{epstopdf}
\usepackage{cancel}
\usepackage{tcolorbox}

\def\xyma{\xymatrix@M.7em}
\def\xymas{\xymatrix@M.1em}

\newcommand{\Comment}[1]{{}}
\definecolor{darkblue}{rgb}{0.15,0.35,0.55}
\definecolor{reddish}{rgb}{0.65, 0.2, 0.2}
\definecolor{darkgreen}{RGB}{50,150,0}
\definecolor{greyish2}{rgb}{.96,.96,.96}
\usepackage[linktocpage=true]{hyperref}
\hypersetup{
colorlinks=true,
citecolor=darkblue,
linkcolor=reddish,
urlcolor=darkblue,
pdfauthor={},
pdftitle={},
pdfsubject={}
}

\usepackage{pgfornament}

\newcommand{\ornamentSep}{\noindent\hfil{{\pgfornament[width=0.3\columnwidth,color=black]{88}}}}

\DeclareFontFamily{OT1}{rsfs10}{}
\DeclareFontShape{OT1}{rsfs10}{m}{n}{ <-> rsfs10 }{}
\DeclareMathAlphabet{\mathscript}{OT1}{rsfs10}{m}{n}

\def\gsim{ \lower .75ex \hbox{$\sim$} \llap{\raise .27ex \hbox{$>$}} }
\def\lsim{ \lower .75ex \hbox{$\sim$} \llap{\raise .27ex \hbox{$<$}} }
\def\be{\begin{equation}}
\def\ee{\end{equation}}
\def\bea{\begin{eqnarray}}
\def\eea{\end{eqnarray}}

\newcommand{\rd}{{\rm d}}
\newcommand{\vp}{\varphi}

\DeclareMathOperator{\E}{e}

\DeclareMathOperator{\arctanh}{arctanh}
\DeclareMathOperator{\sech}{sech}
\DeclareMathOperator{\csch}{csch}

\newcommand{\D}{{\rm d}}

\newcommand{\hypergeom}[2]{
  \mathbin{_{#1}{\sf F}_{#2}} }

\usepackage{latexsym,amsmath,amssymb,epsfig}

\usepackage{tikz}
\usetikzlibrary{decorations}
\pgfdeclaredecoration{complete sines}{initial}
{
    \state{initial}[
        width=+0pt,
        next state=upsine,
        persistent precomputation={\pgfmathsetmacro\matchinglength{
            \pgfdecoratedinputsegmentlength / int(\pgfdecoratedinputsegmentlength/\pgfdecorationsegmentlength)}
            \setlength{\pgfdecorationsegmentlength}{\matchinglength pt}
        }] {}
    \state{upsine}[width=\pgfdecorationsegmentlength,next state=downsine]{
        \pgfpathsine{\pgfpoint{0.25\pgfdecorationsegmentlength}{0.5\pgfdecorationsegmentamplitude}}
        \pgfpathcosine{\pgfpoint{0.25\pgfdecorationsegmentlength}{-0.5\pgfdecorationsegmentamplitude}}
    }
    \state{downsine}[width=\pgfdecorationsegmentlength,next state=upsine]{
        \pgfpathsine{\pgfpoint{0.25\pgfdecorationsegmentlength}{-0.5\pgfdecorationsegmentamplitude}}
        \pgfpathcosine{\pgfpoint{0.25\pgfdecorationsegmentlength}{0.5\pgfdecorationsegmentamplitude}}
}
    \state{final}{}
}

\definecolor{greyish}{rgb}{.90,.90,.90}
\definecolor{greyish2}{rgb}{.96,.96,.96}
\usepackage{xcolor,colortbl}
\usepackage{tcolorbox}

\usepackage[all]{xy}

\usepackage{ytableau}


\numberwithin{equation}{section}

\begin{document}
%
\renewcommand{\thefootnote}{\fnsymbol{footnote}}
\vspace{0truecm}
\thispagestyle{empty}

\begin{center}
{\fontsize{21}{18} \bf Ladder Symmetries of Black Holes:}\\[14pt]
{\fontsize{16}{18} \bf  Implications for Love Numbers and No-Hair Theorems}
\end{center}

\vspace{.15truecm}

\begin{center}
{\fontsize{13}{18}\selectfont
Lam Hui,${}^{\rm a}$\footnote{\texttt{lh399@columbia.edu}} Austin Joyce,${}^{\rm b}$\footnote{\texttt{a.p.joyce@uva.nl}} Riccardo Penco,${}^{\rm c}$\footnote{\texttt{rpenco@andrew.cmu.edu}}\\[4.5pt]
Luca Santoni,${}^{\rm a}$\footnote{\texttt{luca.santoni@columbia.edu}} and
Adam R. Solomon${}^{\rm c}$\footnote{\texttt{adamsolo@andrew.cmu.edu}}
}
\end{center}
\vspace{.4truecm}

 \centerline{{\it ${}^{\rm a}$Center for Theoretical Physics, Department of Physics,}}
 \centerline{{\it Columbia University, New York, NY 10027}} 
 
  \vspace{.3cm}
 
 \centerline{{\it ${}^{\rm b}$Delta-Institute for Theoretical  Physics,}}
 \centerline{{\it University of Amsterdam, Amsterdam, 1098 XH, The Netherlands}}
 
  \vspace{.3cm}

\centerline{{\it ${}^{\rm c}$McWilliams Center for Cosmology, Department of Physics,}}
\centerline{{\it Carnegie Mellon University, Pittsburgh, PA 15213}}
 \vspace{.25cm}

%
%

\vspace{.3cm}
\begin{abstract}
\noindent
It is well known that asymptotically flat black holes in general
relativity have a vanishing static, conservative tidal response. We show that this is a result of linearly realized symmetries governing 
static (spin 0,1,2) 
perturbations around black holes. The symmetries have a geometric origin: in the scalar case, they arise from the (E)AdS isometries of a dimensionally reduced black hole spacetime. Underlying the symmetries is a ladder structure which can be used to construct the full tower of solutions,
and derive their general properties: (1) solutions that decay with
radius spontaneously break the symmetries, and must 
diverge at the horizon; 
(2) solutions regular at the horizon respect the symmetries, and
take the form of a finite polynomial that grows with radius. 
Taken together, these two properties imply that static response coefficients---and in particular Love numbers---vanish. Moreover, property (1) is {consistent with the absence} of black holes with linear (perturbative) hair. We also discuss the manifestation of these symmetries in the effective point particle description of a black hole, showing explicitly that for scalar probes 
the worldline couplings associated with a non-trivial tidal response and scalar hair  must vanish in order for the symmetries to be preserved.

\end{abstract}

\newpage

\setcounter{tocdepth}{2}
\tableofcontents
\newpage
\renewcommand*{\thefootnote}{\arabic{footnote}}
\setcounter{footnote}{0}




\section{Introduction}
\label{intro}

Chandrasekhar famously wrote~\cite{Chandrasekhar:1985kt}: ``the black holes of nature are the most
perfect macroscopic objects there are in the universe: the only
elements in their construction are our concepts of space and time.''
Indeed, the nonlinear vacuum solutions of the Einstein equation,
first discovered a century ago, are elegant in their simplicity,
yet deeply mysterious. Students of black holes learn to cajole them
to give up their secrets by thought experiments: we throw objects at them
and see what comes out; we perturb them and see how
they respond. Of the many remarkable properties of black
holes, we wish to focus on two: the vanishing of the so called Love
numbers, and the absence of hair. 

Consider a black hole in a tidal environment, which could take the
form of an external gravitational field generated by other objects, but
could also be an external scalar or vector field. The situation is
analogous to placing a dielectric under the influence of an external
electric field. The charges inside the material get
displaced, setting up a multipole moment which sources an observable
  response field. 
For black holes, though, such a polarizability, or tidal response, appears to
be absent. Let us be more precise: under the influence of a weak
  static tidal field (spin 0, 1, or 2), black holes have vanishing
Love numbers~\cite{1972ApJ...175..243P,Martel:2005ir,Fang:2005qq,Damour:2009va,Damour:2009vw,Binnington:2009bb,Kol:2011vg,Landry:2015cva,Landry:2015zfa,Gurlebeck:2015xpa,Porto:2016pyg,Poisson:2020mdi,LeTiec:2020spy,LeTiec:2020bos,Chia:2020yla,Goldberger:2020fot,Hui:2020xxx,Charalambous:2021mea}. 
The Love numbers quantify the tidal response: 
suppose the external tidal field goes as $r^\ell$ at large distance
$r$ ($\ell$ is the multipole of interest), an object such as a star
that gets tidally deformed would source a response field
that goes as $1/r^{\ell+1}$ far away.\footnote{The precise exponent depends on the choice of field one
  focuses on. For instance, for the gravitational field, while the metric
fluctuations have $r^\ell$ and $1/r^{\ell+1}$ asymptotics at large $r$,
the Weyl tensor asymptotes to $r^{\ell-2}$ and $1/r^{\ell+3}$.
} The coefficient of this $1/r^{\ell+1}$
tail, normalized by the amplitude of the tidal field, 
is the Love number, or rather Love numbers (or Love tensors
even), for there are in general different types of tidal fields one
could turn on.\footnote{There is some discussion in the literature on exactly how to define
   Love numbers~\cite{Martel:2005ir,Damour:2009va,Gralla:2017djj,LeTiec:2020spy,LeTiec:2020bos}. Here, we
  follow the definition laid out in
  \cite{Goldberger:2004jt,Porto:2007qi,Kol:2011vg,Goldberger:2020fot,Hui:2020xxx,Charalambous:2021mea}.
  The Love numbers are identified with the coefficients of operators
  in the point-particle action involving the square of the tidal
  field (in the case of spin 2, it is the square of the Weyl tensor,
  suitably contracted). As such, the definition is explicitly gauge
  invariant, and solutions to the Teukolsky equation can be related
  directly to the Love numbers. We rely in particular on the matching
  computations in~\cite{Hui:2020xxx,Charalambous:2021mea}. This definition
  differentiates non-dissipative response (what we are interested in) from
  the dissipative one, relevant especially for rotating black holes
  \cite{Porto:2007qi,LeTiec:2020spy,Chia:2020yla,Goldberger:2020fot,LeTiec:2020bos,Charalambous:2021mea}.
}

For a weak static tidal field, the problem reduces to that of studying
static linear perturbations around the object of interest, be it a star or a
black hole. At a large distance from the object, the perturbation, 
suitably defined (let's call it $\phi$), obeys
\begin{equation}
\label{poisson}
\nabla^2 \phi = 0 \, ,
\end{equation}
{which admits two {linearly independent} solutions at large $r$, one with a growing $r^\ell$ profile and one with a decaying $1/r^{\ell+1}$ profile at multipole order
$\ell$.}\footnote{{We will therefore use the term ``decaying branch''
  to refer to the purely decaying solution that goes as $1/r^{\ell+1}$
  at large $r$, and refer to the solution with pure $r^\ell$
  asymptotics (no $1/r^{\ell+1}$ contribution) as the ``growing
  branch''. A general solution is a linear superposition of the two
  branches.}}
For a star, the growing (tidal) solution is generally accompanied by the
decaying tail, signifying deformability. For a black hole, the
growing solution is regular at the horizon while the decaying one is
not (and therefore discarded). These two facts combined tell us its
Love numbers vanish.\footnote{More details about the growing solution are needed, in
  particular that as a power series in $r$, only certain powers appear, as will be explained below.
} 

The second fact, that the decaying solution diverges at the horizon,
ought to remind us of no-hair theorems.
No-hair theorems tell us we would inevitably fail if we attempt to endow
black holes with extra features, in other words, hair that is
static and falls off with distance. 
{As we will discuss below, from the perspective of the second order linear
  differential equation, just as $\phi$ has two possible asymptotics at
  large $r$ (i.e. $r^\ell$ and $1/r^{\ell+1}$), it also has two
  possible behaviors close to the horizon (i.e.~regular and singular). 
  Hair---if it exists---would correspond to a solution that is purely decaying at
  infinity and is also regular at the horizon of the black hole. As
  such, having hair would be somewhat surprising, 
  because our generic expectation is that a purely decaying solution at infinity would map to an admixture of regular and singular modes near the horizon. Indeed, we will see that we can use symmetry arguments to confirm this expectation.
Of course, the full no-hair theorems are
stronger and more nontrivial} than the version we focus on in this paper: they apply even
when some nonlinear interactions are present, and/or when the metric is not
a solution to the Einstein equation (but crucially must have a
horizon)~\cite{Israel:1967wq,Carter:1968rr,Carter:1971zc,Wald:1971iw,Hartle:1971qq,Bekenstein:1971hc,Fackerell:1972hg,Price:1972pw,Bekenstein:1995un,Hui:2012qt}.\footnote{We would be remiss not to mention exceptions: they generally involve
  hair that is not static, or does not fall off at infinity, or is
  non-minimally coupled to gravity, or has special interactions \cite{Jacobson:1999vr,Lee:1991vy,Weinberg:2001gc,Alexander:2009tp,Sotiriou:2013qea,Babichev:2013cya,Sotiriou:2014pfa,Herdeiro:2014goa,Silva:2017uqg,Wong:2019yoc,Clough:2019jpm,Hui:2019aqm}.}
Nonetheless, our limited version---for linear, scalar/vector/tensor
hair around a Schwarzschild or Kerr black hole---is a manifestation of
these theorems, and it connects naturally with the phenomenon of tidal response.\footnote{In this paper, by
  ``hair'' we mean any property or quantity that a generic
  gravitational object (such as a star or a planet) can have but a general
  relativistic black hole does not have (i.e., any property or quantity
  beyond the standard three: mass, spin and electromagnetic charge).
  We do not distinguish between primary versus secondary hair
  \cite{Herdeiro:2015waa}. In this sense, black holes in a theory with
the scalar-Gauss--Bonnet coupling have scalar hair
\cite{Sotiriou:2013qea} (see also Appendix~\ref{app:GB}).
A planet with a complex landscape (i.e., mountains and valleys) has
complicated multipole moments in its exterior gravitational field,
beyond those of a Kerr black hole~\cite{Hansen:1974zz}, and thus also has {``gravitational multipole"} hair.
}

{In contrast to linear hair, the vanishing of Love numbers is
  more of a surprise: our natural expectation would again be that the
  solution that is regular at the horizon will be an admixture of an
  applied tidal field at infinity ($r^\ell$) along with its induced
  response ($1/r^{\ell+1}$). The surprise is that the solution to the relevant linear differential equation regular at the horizon is also pure tidal field at infinity---meaning it has a single fall-off in two asymptotic regimes. This is in contrast to the situation for generic objects, where boundary conditions lead to both a tidal field and its response at infinity, implying a static response. The special status of black holes in the space of objects 
 begs for a symmetry explanation, and in fact one exists.} 
 We will work out the symmetries governing static, linear
perturbations of general spin around Schwarzschild or Kerr black holes. 
Underlying them is a ladder structure that connects
solutions at different~$\ell$s. The symmetries give rise to conserved
quantities which can be used to connect asymptotic behaviors at the
horizon and at infinity. Interestingly, the growing branch linearly realizes the symmetries while the decaying one does not. 
This gives us a richer perspective on the choice of one branch over
another: the boundary conditions that differentiate between a star and a
black hole are seen to pick out states that make or break the ladder
symmetries. In the scalar case at least, the symmetries have an identifiable
geometric origin: they follow from the isometries of an Euclidean
anti-de Sitter space of a suitably rescaled, dimensionally reduced
black hole spacetime. Along the way, we will uncover a second ladder
structure, one that connects perturbations of different spin.


The symmetry understanding is interesting from an infrared
perspective. A far away observer, if they detect a $\phi \sim
1/r^{\ell+1}$ profile, would conclude that the object of interest has some
sort of (multipolar) charge that sources such a profile. That is, they
would model it by adding a source term to eq.~\eqref{poisson}, localized on the
object and proportional to the charge.\footnote{The approach of modeling distant behavior by adding terms
  to the point particle action is a standard effective field theory
  technique (see e.g.,~\cite{Goldberger:2004jt}).
}
A black hole cannot have
such a charge, or hair. As we will see, the ladder symmetries have a non-trivial large
distance limit, telling us that a charge-less object like the black
hole is an object with enhanced symmetries. Likewise, if the far away
observer finds a $1/r^{\ell+1}$
tail when an $r^\ell$ tidal field is applied, they would conclude the object of interest has a non-trivial
static response. That is, they would model it by adding another source term to
eq.~\eqref{poisson}, localized at the position of the object and proportional to
the tidal field. The proportionality constant is the
Love number. In the effective worldline description of the system, these couplings break the infrared manifestations of the symmetries, so that the point in parameter space where these couplings vanish (as is appropriate to model a black hole) is an enhanced symmetry point.

A few papers are particularly relevant to our investigation.
One is an interesting recent paper by Charalambous, Dubovsky, and Ivanov
\cite{Charalambous:2021kcz}. They pointed out a different set of
symmetries that are also relevant for the vanishing of the black hole
Love numbers. Their symmetry transformation involves time, while ours
is purely static. The connection between the two is an interesting
question which we hope to explore in the future.
Another influential paper is by Compton and Morrison \cite{Compton:2020cjx}. 
The ladder structure we find for black holes mirrors the one they
discovered for de Sitter space. As we will explain, there is a simple
reason: at least for the scalar case, the relevant metric, suitably reduced and
rescaled, can be analytically continued to de Sitter. Our strategy for
identifying conserved charges follows theirs.
The use of conformal Killing vectors to
relate different scalar theories (see Appendix \ref{geometry}) 
was pointed out by Cardoso, Houri, and Kimura \cite{Cardoso:2017qmj}.

The discussion is organized as follows.
We start with the simplest case in Section \ref{ladderSchw}:  
a massless scalar in a Schwarzschild background. 
All the essential features
that govern higher spin perturbations in 
a Kerr background can be found in this case. 
Section \ref{ladderKerr} covers a scalar in the Kerr spacetime, and 
Section \ref{ladderSpin} extends the argument to vector and tensor 
perturbations. In Section \ref{IRsymmetries}, we discuss the 
infrared limit of the ladder symmetries.
We conclude in Section \ref{discuss}, commenting on 
the extension to finite frequency and general number of dimensions.
The Appendices contain a number of useful results: 
Appendix \ref{geometry} on the geometric origin of the ladder
symmetries, \ref{app:PT} on the P\"oschl--Teller potential and its
relation to our problem,
\ref{ladderReggeWheeler} and \ref{ladderTeukolsky} on the
ladder structures in the Regge--Wheeler and Teukolsky equations
respectively, 
\ref{SUSY} on the supersymmetric structure of the ladder symmetries,
\ref{hyperg} on certain hypergeometric identities helpful for
revealing the ladders,
and \ref{app:GB} on black hole hair and tidal response for a scalar
coupled to the Gauss--Bonnet curvature invariant\hspace{.02cm}\raisebox{-0.05ex}{\includegraphics[scale=.0045]{cow.pdf}}

\newpage
\section{Ladder in Schwarzschild}
\label{ladderSchw}

The simplest manifestation of the ladder symmetry structure we wish to
describe is in the equation for a massless scalar propagating in a
fixed Schwarzschild background.
The Schwarzschild metric is
\be
\rd s^2 = -{\Delta \over r^2} \rd t^2 + {r^2 \over \Delta} \rd r^2 + r^2 (\rd \theta^2 + \sin^2 \theta  \, \rd\varphi^2)\,,
\ee
where $\Delta \equiv r (r - r_s)$ and $r_s$ is the Schwarzschild radius. 
A static, massless scalar $\phi$ in this background satisfies the equation of motion
\be
\label{phieom}
\partial_r (\Delta \partial_r \phi_\ell) - \ell (\ell + 1) \phi_\ell = 0 \, ,
\ee
where we have decomposed $\phi$ in spherical harmonics, i.e., $\phi_\ell$ is the field amplitude associated with a particular $Y_{\ell m}(\theta,\vp)$, with $\ell$ the angular momentum quantum number
(the equation is independent of the magnetic quantum number $m$, and therefore we will suppress this index in what follows).
It will prove convenient to multiply eq.~\eqref{phieom} by $-\Delta$, rewriting the
equation of motion as:
\be
H_\ell \phi_\ell = 0 \,, \qquad {\rm where} \qquad H_\ell \equiv -\Delta \big(\partial_r (\Delta \partial_r) - \ell (\ell + 1)\big) \, ,
\label{eq:schscalarEOM}
\ee
which defines a ``Hamiltonian," $H_\ell$, where derivatives act on everything to their right.
For large $r$, it is easy to see that the two linearly independent solutions to eq.~\eqref{phieom} are $\phi_\ell \sim r^\ell$ (growing) or $1/r^{\ell+1}$ (decaying). For $r \rightarrow r_s$, $\phi_\ell$ goes either as a constant or a logarithm,
$\ln[(r - r_s)/r_s]$. The tidal response of a black hole is determined by the ratio of the decaying and growing modes at infinity, calculated under the assumption that the scalar remains regular at the horizon. The key task is therefore to connect the growing/decaying {profile at infinity} to the horizon asymptotics.

Let us define the following operators:
\be
\begin{aligned}
D^+_\ell &\equiv - \Delta \partial_r +  \frac{\ell + 1}{2}(r_s - 2r) \, , \\
D^-_\ell &\equiv \Delta \partial_r + \frac{\ell}{2} (r_s - 2r) \, .
\end{aligned}
\label{eq:schwarzschildscalarladders}
\ee
The geometric origin of these operators is discussed in Appendix
\ref{geometry}. An explicit calculation shows that they satisfy the
following relations:\footnote{Though it is not important to our
  argument, it is worth noting that if one deforms $D^\pm$ 
  by introducing a phase that catalogues the $\ell$ value as:
\be
\begin{aligned}
L_\pm \equiv \frac{2}{r_s}e^{\pm i\alpha}\left(\mp\Delta\partial_r+\frac{1}{2}\Delta'(r) i\partial_\alpha\right)\,,\qquad\qquad\quad L_0 \equiv -i\partial_\alpha\,,
\end{aligned}
\ee
then $L_\pm$ and $L_0$ satisfy an SL$(2)$ algebra: $[L_\pm,L_0] = \mp L_\pm$, $[L_+,L_-] =-2L_0$.
}
\be
\begin{aligned}
 &D^-_{\ell+1} D^+_\ell   - D^+_{\ell-1} D^-_\ell = \frac{(2\ell + 1)r_s^2}{4} \,,  \\
 &H_\ell = D^-_{\ell + 1} D^+_\ell  - {(\ell +1)^2 r_s^2\over 4} = D^+_{\ell-1} D^-_\ell - {\ell^2 r_s^2\over 4}\,.
\end{aligned}
\label{eq:schladderalg}
\ee
The $D^\pm_\ell$ operators are raising and lowering operators
in the sense that their algebra with the Hamiltonian is\footnote{In this sense, the ladder structure that we are discussing is slightly different from the more familiar ladder structure in, for example, the harmonic oscillator. Here the ladder operators map a solution of a given Hamiltonian---indexed by $\ell$---to a corresponding eigenstate of a {\it different} Hamiltonian, indexed by $\ell \pm 1$. This is in contrast to the harmonic oscillator where ladder operators map between different states of the same Hamiltonian.}
\be
\label{HD}
H_{\ell + 1} D^+_\ell = D^+_\ell H_\ell \, , \qquad\quad
H_{\ell-1} D^-_{\ell } = D^-_{\ell} H_{\ell} \, .
\ee
Using this structure,
from a solution at level $\ell$, i.e., $\phi_\ell$ such that
$H_\ell \phi_\ell = 0$, we can raise to  a level $\ell+1$ solution. That is, 
$D^+_\ell \phi_\ell$ solves $H_{\ell + 1} (D^+_\ell \phi_\ell) = 0$. 
Similarly, we can lower to a level $\ell-1$ solution, i.e.,
$D^-_\ell \phi_\ell$ solves $H_{\ell-1} (D^-_\ell \phi_\ell) = 0$.
Also, both $D^-_{\ell+1} D^+_{\ell} \phi_\ell$ and $D^+_{\ell-1}
D^-_\ell \phi_\ell$ reproduce $\phi_\ell$ (up to a constant
normalization), provided $H_\ell \phi_\ell = 0$.

From this algebraic structure
a whole tower of solutions falls into our lap immediately. First, the shift symmetry of the scalar
guarantees that $\phi_0 =$ constant
is a good $\ell = 0$ solution; then applying
$D^+_{\ell-1} \cdots D^+_1 D^+_0 \phi_0$ gives
us the corresponding solution at level $\ell$. From the form 
of the raising operator, one can see the level $\ell$ solution
is a polynomial in $r$ of the form $\phi_\ell \sim 1 + r + \cdots + r^\ell$. 
Such a finite polynomial is manifestly regular at the horizon, $r = r_s$. 
{This solution is of the growing type:} going as $r^\ell$ at large $r$.

It is not difficult to write down the precise decaying solution for
$\ell = 0$  and raise it to arbitrary $\ell$.\footnote{The exact $\ell=0$ decaying solution
is ${\rm ln\,}[(r-r_s)/r]$.
}
Instead we take a different tack: let us delve deeper into the
symmetries behind the ladder structure. Here, we follow the strategy of
\cite{Compton:2020cjx} who pointed out an analogous ladder
in de Sitter space.\footnote{In fact, as we discuss in Appendix~\ref{geometry}, one can change variables to cast the static equation of motion~\eqref{phieom} as the equation for a (massive) scalar in a three-dimensional Euclidean anti-de Sitter geometry. From this perspective, the vanishing of the static response is related by analytic continuation to the transparency of odd-dimensional de Sitter spaces.}

We can use the algebraic structure to construct conserved quantities for each $\ell$. At the $\ell=0$ level, it is easy to see:
\be
\label{Q0def}
Q_0 \equiv \Delta \partial_r\,\quad{\rm satisfies}\quad [Q_0 , H_0] = 0. 
\ee
Thus $Q_0$ generates a symmetry in the $\ell = 0$ sector.
Climbing the ladder, we can construct $Q_1 \equiv D^+_0 Q_0 D^-_1$
which satisfies $[Q_1, H_1] = 0$, i.e.,
$D^+_0 Q_0 D^-_1 H_1 - H_1 D^+_0 Q_0 D^-_1
= D^+_0 [Q_0, H_0] D^-_1 = 0$.
Proceeding inductively, we have:
\be
\label{Qelldef}
Q_\ell \equiv D^+_{\ell-1} Q_{\ell-1} D^-_\ell \,\quad{\rm which~satisfies}\quad [Q_\ell, H_\ell] = 0\,.
\ee
It can be shown $\delta \phi_\ell = Q_\ell \phi_\ell$ is a symmetry of
the action, from which a corresponding Noether current can be
derived.\footnote{Strictly speaking, there should be a
dimensionful factor relating $\delta \phi_\ell$
and $Q_\ell \phi_\ell$, which we keep implicit.
}
In the static limit, current conservation takes the form
$\partial_r J^{\, r}_\ell(r) =0$, i.e., the $r$ component of the
Noether current is a {\it conserved} quantity, in the sense of being 
{\it $r$-independent}. For this reason, we will  refer to it as ``charge'' in what follows. 

It is simpler to identify the conserved quantities by inspection. Equation (\ref{phieom}) tells us, for $\ell=0$:
\be
 P_0 \equiv \Delta\partial_r \phi_0\,\quad{\rm is~conserved:}\quad\partial_r P_0 = 0 \,,
\ee
where $\phi_0$ is now an arbitrary solution of the $\ell =0$ equation. 
Conserved quantities at higher $\ell$ can be obtained by climbing the ladder:
\be
\label{higherQ}
\partial_r P_\ell = 0 \,,\quad {\rm where} \quad P_\ell \equiv \Delta\partial_r (D^-_1 D^-_2 \cdots D^-_{\ell} \phi_\ell) \, .
\ee
The philosophy is that, starting from any solution at level $\ell$, one can descend to $\ell=0$ by
a series of lowering operations, and then invoke the conservation of
$P_0$ to see that $P_\ell$ is conserved. The Noether current
associated to the symmetry  $\delta \phi_\ell = Q_\ell \phi_\ell$
turns out to be actually $J^{\, r}_\ell = (P_\ell)^2$, but clearly
this implies that $P_\ell$ is also conserved.\footnote{Notice that $P_\ell$ and $\delta\phi_\ell$ happen to
  coincide for $\ell =0$, but are different from each other for
  non-zero values of $\ell$.}
We will sometimes abuse the
terminology and refer to $P_\ell$ itself as the conserved charge.

The utility of the conserved quantities 
$P_\ell$ is that they can be used to deduce properties of the solutions to~\eqref{phieom}, without knowing their precise form. In particular this allows us to understand how solutions with particular behaviors in the two asymptotic regions (near the horizon and near $r=\infty$) connect to each other without explicitly solving the differential equation.
Consider the decaying solutions, which go like
$\phi_\ell \sim 1/r^{\ell+1}$ at large $r$. Each $D^-$ operator essentially
increases the power in $r$ by $1$ at large $r$. Thus $D^-_1 \cdots D^-_{\ell} (1/r^{\ell+1})$
scales as $1/r$ at large $r$, so that its corresponding $P_\ell$ is nonzero. Near the horizon, this solution must have the same $P_\ell$, since it is conserved as a function of $r$. Of the two possible horizon asymptotics, the constant solution yields $P_\ell = 0$ while the logarithm, $\ln[(r - r_s)/r_s]$, yields a non-vanishing $P_\ell$ as $r \rightarrow r_s$. Thus, conservation of $P_\ell$ tells us the decaying  solutions must diverge logarithmically at the horizon.\footnote{The solution that is purely decaying mode at infinity $\sim 1/r^{\ell+1}$ is an admixture of the constant and logarithmic fall-offs near the horizon, but the important point is that it is not regular at the horizon.}

We already know that the growing branch of solutions, defined by $\phi_\ell = D^+_{\ell-1} \cdots D^+_0 \phi_0$, with a constant $\phi_0$, is regular at the horizon (by virtue of being a finite polynomial). But we can see this by charge conservation, too. The series of
lowering operators in $P_\ell$ merely reverts $\phi_\ell$ back to the
constant $\phi_0$ (up to normalization) which implies that $P_\ell =
0$ for the growing branch. Conservation of $P_\ell$ thus tells us this solution, when
extended to the horizon, must approach a constant rather than the
divergent logarithm.\footnote{Interestingly, charge conservation implies that the solution relevant for the computation of Love numbers has a single fall-off when expanded in each of the asymptotic regions. This is closely related to the phenomenon of reflection-less transmission in quantum mechanics, which we elaborate on in Appendix~\ref{app:PT}.}

It is worth emphasizing the vanishing of the Love number follows from {\it two} facts: (1) the purely decaying solution ($\sim 1/r^{\ell + 1}$ at large $r$) is divergent at the horizon, and (2) the solution that is regular at the horizon is a finite polynomial going
as $1 + r + \cdots + r^\ell$. The second fact by itself is not
sufficient: it does not forbid us from adding to it a decaying tail
that goes as $1/r^{\ell+1}$.\footnote{In order to infer that Love
  numbers vanish, in addition to the polynomial solution, we need to
  know that the other solution diverges at the horizon. This can
  alternatively be inferred from analysis of the differential equation
  in the near-horizon regime. However, to make statements about black
  hole hair we need to be able to connect the solution that goes as
  $1/r^{\ell+1}$ at infinity to these near-horizon solutions, which
  fact (1) achieves.} 
A black hole, by the demand of regularity at the horizon, cannot have
this tail---implying that its static response vanishes, meaning zero
Love number. 

Fact (1) is also the content of the no-hair theorem. Indeed, Bekenstein showed that 
a black hole cannot sustain static, scalar hair that decays at infinity \cite{Bekenstein:1971hc}.
His celebrated proof makes minimal assumptions about the spacetime{---relying essentially only on the existence of a horizon---and is thus more general}. Our derivation shows that, {in the limited case of massless scalar a Schwarzschild background, the absence of hair can also be deduced from a symmetry argument based on the very ladder symmetries that imply the vanishing of static response.}

 It is worth stressing that Bekenstein's no-hair theorem applies even
 if the scalar has a mass, or some non-trivial interactions. In such a case,
 our ladder symmetries---for a linear, massless scalar---cannot be
 used to justify the corresponding no-hair property (see further discussion in
Sections \ref{IRsymmetries} and \ref{discuss}).
Our seemingly narrow focus on a linear,
 massless scalar is motivated by the eventual goal of studying linearized metric
 perturbations around black holes: as we will see
 in Section \ref{ladderSpin}, a different ladder structure allows us
 to infer properties of the linear spin-2 perturbations directly from those of the linear,
 massless scalar.\footnote{As an alternative line of reasoning, one could take Bekenstein's no-hair
theorem as a given, and use it to justify fact (1). Our main point
is that both facts (1) and (2), for a linear, massless scalar, can be
deduced from the ladder symmetries.
}

A pleasing feature of the story laid out so far is that one can also motivate the
disposal of the decaying solution by {\it symmetry}, rather than by
the requirement of {\it regularity}. For instance, at the
$\ell = 0$ level, it is simple to check that the symmetry transformation
$\delta \phi_ 0 \equiv Q_0 \phi_0$ vanishes if $\phi_0$ is a constant
(the growing solution), and is non-zero if $\phi_0$ goes as $1/r$ at
large $r$ (the decaying solution). The same pattern can be established
for the higher $\ell$'s by ascending the ladder. That is, 
$\delta \phi_\ell \equiv Q_\ell \phi_\ell$ vanishes for the
growing branch (thus respecting the symmetry)
and is non-zero for the decaying branch (thus spontaneously breaking
the symmetry). The symmetry perspective is
helpful especially when viewing the black hole
from far away---such an observer might 
not even be aware of the existence of a horizon
(let alone regularity on it), but could
determine whether the field configuration at large $r$ 
is symmetric. This is discussed further in Section~\ref{IRsymmetries}.

As an aside, it is worth 
stressing there are two different kinds of symmetries at
work. One mixes $\ell$ levels, which we call a {\it vertical} symmetry. It acts on the fields as
\be
\begin{aligned}
\label{pairDs}
\delta \phi_\ell = D^+_{\ell-1} \phi_{\ell-1}\,, \qquad  \qquad
\delta \phi_{\ell-1} = -D^-_\ell \phi_{\ell} \, .
\end{aligned}
\ee
This is a symmetry of the action, and has a geometric origin
(see Appendix \ref{geometry}). 
Its existence is responsible for the ladder structure.
The other symmetry---associated with 
the conservation of $P_\ell$---stays within the same level (acts on a given rung of the ladder). We call this a
{\it horizontal} symmetry: $\delta \phi_\ell = Q_\ell
\phi_\ell$. It owes its existence to the horizontal symmetry at
$\ell=0$, which can then be bootstrapped to higher $\ell$'s by climbing the ladder.

It is also worth mentioning that we have specialized to $D=4$, but the equation of motion for a scalar in general dimension can be cast in the form~\eqref{phieom} with $\ell$ replaced with $\hat\ell \equiv \ell/(D-3)$. All of the same ladder structure is present in general dimensions, but only integer $\hat\ell$ can be connected to $\hat\ell = 0$ via the ladder, so only these values have conserved quantities. Indeed, precisely these values have vanishing Love numbers in general dimension~\cite{Kol:2011vg,Hui:2020xxx}.

We next extend the discussion to a scalar field in the Kerr geometry, before treating the general case.

\newpage
\section{Ladder in Kerr}
\label{ladderKerr}

The algebraic ladder structure that we have discussed has a direct analogue in a Kerr background.
The Kerr line element in Boyer--Lindquist (BL) coordinates is:
\be
\begin{aligned}
\rd s^2 &= -\frac{\Delta}{\rho^2}\left(\rd
  t-a\sin^2\theta\,\rd\vp\right)^2+\frac{\rho^2}{\Delta}\rd r^2+\rho^2
\rd\theta^2+\frac{\sin^2\theta}{\rho^2}\left(a\,\rd
  t-(r^2+a^2)\rd\vp\right)^2  \\
 &= -{\rho^2 - r_s r \over \rho^2} \rd t^2  - {2a r_s r {\,\rm sin}^2\theta
  \over \rho^2} \rd t \rd \varphi + {\rho^2 \over \Delta} \rd r^2 + \rho^2
\rd \theta^2 + { (r^2 + a^2)^2 - a^2\Delta {\,\rm sin}^2\theta \over
  \rho^2} {\,\rm sin}^2\theta \rd \varphi^2
,
\end{aligned}
\ee
where we have defined the quantities
\be
\rho^2 \equiv r^2 +a^2 \cos^2\theta \, , \qquad\quad
\Delta \equiv r^2 - rr_s + a^2 \, , 
\ee
where also $\Delta = (r-r_+)(r-r_-)$ with $r_\pm \equiv r_s/2 \pm \sqrt{(r_s/2)^2 - a^2}$ being the locations of the inner and outer horizons. 
It is useful to note the relations: $r_+ r_- = a^2$, $a^2 + r_+^2 =
r_s r_+$
and $g_{t\varphi} {}^2 - g_{tt} g_{\varphi\varphi} = \Delta {\,\rm
  sin}^2\theta$. 
The equation of motion for a static, massless scalar in a Kerr background is,
after decomposing in spherical harmonics $Y_{\ell m}$:
\be
\label{eomKerr}
\partial_r (\Delta \partial_r \phi_\ell) + \frac{a^2 m^2}{\Delta} \phi_\ell - \ell(\ell+1)\phi_\ell = 0\,,
\ee
where the magnetic quantum number, $m$, dependence of $\phi_\ell$ is
again suppressed. 
Note that $Y_{\ell m} \propto \E^{im \varphi}$. 
It is useful to make the following definitions:
\be
\begin{aligned}
\label{phipsi}
 z &\equiv r - r_- \,, &  \qquad z_k &  \equiv r_+ - r_- \, \\
 q &\equiv \frac{am}{z_k}\,, &  \qquad
\phi_\ell &\equiv \E^{i q \ln [(z-z_k)/z]} \psi_\ell \, .
\end{aligned}
\ee
Note that $\partial_r \E^{i q \ln [(z-z_k)/z]} =  (iam/\Delta) \E^{i q
  \ln [(z-z_k)/z]}$, $\Delta = z(z - z_k)$, and $2r - r_s = 2z -
z_k$. 
After these changes of variables, the scalar equation of motion can thus be recast as:
\be
\partial_z ( \Delta \partial_z \psi_\ell) + 2 iq z_k \partial_z \psi_\ell -
\ell(\ell + 1)\psi_\ell = 0 \, .
\label{eq:kerreomz}
\ee
As before, it is convenient to multiply this equation by $-\Delta$ and rewrite:
\be
H_\ell \psi_\ell = 0 \,  , \quad{\rm with}\quad \, \, H_\ell \equiv -\Delta\big(\partial_z (\Delta\partial_z) + 2iq z_k \partial_z - \ell(\ell +1)\big) \, .
\ee
At large $z$, the growing/decaying solutions go as $z^\ell$ or
$1/z^{\ell+1}$, respectively. Close to the horizon ($z \rightarrow z_k$), 
the two asymptotic solutions are $\psi_\ell \sim$ constant,
and $\psi_\ell \sim \E^{-2iq\ln[(z - z_k)/z_k]}$. 

The equation~\eqref{eq:kerreomz} has a similar ladder structure to the Schwarzschild scalar. It can be discerned by defining: 
\be
\begin{aligned}
\label{DKerr}
 D^+_\ell &\equiv -\Delta \partial_z + \frac{(\ell+1)}{2}(z_k - 2 z) -  iq z_k\, , \\
 D^-_\ell &\equiv \Delta \partial_z +  \frac{\ell}{2}(z_k - 2z) + iq z_k \, .
\end{aligned}
\ee
These operators obey very similar relations to~\eqref{eq:schladderalg}: 
\be
\begin{aligned}
 D^-_{\ell+1} D^+_\ell   - D^+_{\ell-1} D^-_\ell &= \frac{(2\ell + 1)z_k^2}{4} \, ,  \\
H_\ell = D^-_{\ell + 1} D^+_\ell  - \left(\frac{(\ell +1)^2}{4} + q^2\right)z_k^2 &= D^+_{\ell-1} D^-_\ell - \left(\frac{\ell^2}{4} + q^2\right) z_k^2 \, .
\end{aligned}
\ee
In addition, exactly the same ``commutators'' as in eq.~\eqref{HD} apply here, from which
the interpretation of $D^\pm$ as raising and lowering operators
follows. Note that the ladder structure extends in $\ell$ but not in
$m$. It is possible to formally climb to any $\ell$, including $\ell <
\lvert m\rvert$. 
Equation~\eqref{eomKerr} does not by itself forbid this. 
It is the connection to $Y_{\ell m}$ that ultimately limits $\lvert
m\rvert \le \ell$.\footnote{
For instance, suppose one is interested in $\psi_1$ for $m=1$. This
can be constructed by climbing from $\psi_0$, {\it for the same} $m=1$. 
The latter is not physical (when viewed in conjunction with $Y_{\ell m}$), but is a perfectly good solution to
equation \eqref{eq:kerreomz} for $\ell=0$ and $m=1$. {That the
raising/lowering operator can be used in this way might seem surprising. But it should be borne in mind
that a raising/lowering operator acting in isolation is {\it not} a
symmetry (as can be seen from eq. \ref{HD}). What is a symmetry (and a
symmetry at the level of the action) is one where the operators act in
concert, such as in eq. (\ref{pairDs}), (\ref{verticalSymmAll}) or
(\ref{QsKerr}).}}


The introduction of $\psi_\ell$ to replace $\phi_\ell$ in eq.~\eqref{phipsi} is useful not only for making the equation resemble its Schwarzschild counterpart, but also for thinking about horizon regularity. The logarithmic phase that relates $\phi_\ell$ and
$\psi_\ell$ is motivated by the azimuthal coordinate that was used originally by Kerr~\cite{Kerr:1963ud}:
$\varphi_{\rm Kerr} = \varphi + (a/z_k) \ln[(z-z_k)/z]$. The point is
that close to the horizon $z \rightarrow z_k$, the $\psi_\ell \sim$ constant solution is regular while the
$\psi_\ell \sim \E^{-2iq\ln[(z - z_k)/z_k]}$ solution is not \cite{1972ApJ...175..243P}.\footnote{
This can be checked by computing $T_{\mu\nu} u^\mu u^\nu$, where
$T_{\mu\nu}$ is the energy-momentum tensor of the scalar $\phi$, and $u^\mu$
is the 4-velocity of a freely falling observer.
For an observer freely falling from rest at infinity, with zero
angular momentum, $u^\mu = \rho^{-2} (\Sigma^2 / \Delta, -
\sqrt{\Sigma^2 - \Delta\rho^2}, 0, ar_s r/\Delta)$, where $\Sigma^2$ 
is defined by $g_{\varphi\varphi} \equiv \Sigma^2
{\,\rm sin}^2\theta/\rho^2$. 
It can be shown $T_{\mu\nu} u^\mu u^\nu$ is finite at the horizon for
the $\psi_\ell \sim$ constant solution but not for the $\psi_\ell \sim \E^{-2iq\ln[(z - z_k)/z_k]}$ solution.
}
Moreover, according to calculations by
\cite{Chia:2020yla,Goldberger:2020fot,Charalambous:2021mea}, the
static Love number can be inferred by examining the power series for
$\psi_\ell$: absence of a $1/r^{\ell+1}$ term in $\psi_\ell$ is 
tantamount to a vanishing (conservative) Love number.

Returning to the ladder structure: just as in the Schwarzschild case we can define a set of horizontal symmetries that act only at a fixed $\ell$. The symmetry transformation is $\delta\psi_\ell = Q_\ell \psi_\ell$ with $Q_\ell$ given by
\be
\label{QsKerr}
\begin{aligned}
 Q_0 &\equiv \Delta\partial_z \,, & \qquad\qquad [Q_0, H_0] &= 0  \, ,
 \\
 Q_\ell &\equiv 
D^+_{\ell-1} Q_{\ell-1} D^-_\ell\,, &  [Q_\ell, H_\ell] &= 0\,. 
\end{aligned}
\ee
The corresponding conserved charge is, as before, most easily obtained by inspection, starting from $\ell=0$:
\be
\label{Q0Kerr}
P_0 \equiv 
(\Delta\partial_z + 2iq z_k)\psi_0 \quad{\rm is~conserved:}\quad\partial_z P_0 = 0\,.
\ee
Higher $\ell$ charges, $P_\ell$, are found by climbing the ladder:
\be
\partial_z P_\ell = 0 \,, \quad {\rm where} \quad P_\ell \equiv
(\Delta\partial_z + 2iq z_k) D^-_1 \cdots D^-_\ell \psi_\ell \, ,
\ee
for any level-$\ell$ solution $\psi_\ell$.

The growing branch of solutions can again be raised from
the $\psi_0 = {\rm constant}$ solution: 
$\psi_\ell  = D^+_{\ell-1} \cdots D^+_0 \psi_0$ is the corresponding level $\ell$ solution.
It is manifestly a finite polynomial $\sim 1 + \cdots + z^\ell$
(from the form of $D^+$), and is thus
regular at the horizon. 

The conserved charge $P_\ell$ evaluated for the
so-defined growing branch is non-zero, unlike in the
Schwarzschild case:
\bea
\label{Pellexp}
P_\ell = {z_k^{2\ell + 1} \over 4^\ell} {2 i q} \left(1 + {4
    q^2}\right) ... \left(\ell^2 +  {4 q^2}\right) 
= {z_k^{2\ell + 1} \over 4^\ell} {i \over \pi} {\,\rm sinh}(2\pi q) \,
|\Gamma (1 + \ell + 2iq) |^2 \, ,
\eea
where we have chosen $\psi_0 = 1$.\footnote{\label{matching}This can be shown by noting that
$P_\ell = (\Delta\partial_z + 2iq z_k) D^-_1 \cdots D^-_\ell
\psi_\ell = (\Delta\partial_z + 2iq z_k) D^-_1 \cdots (D^-_\ell
D^+_{\ell-1}) D^+_{\ell-2} \cdots D^+_0 \psi_0$. 
We highlight by parentheses $(D^-_\ell D^+_{\ell-1})$ and observe
that to its right is $\psi_{\ell-1}$ which solves $H_{\ell-1}
\psi_{\ell-1} = 0$, which tells us $[D^-_\ell D^+_{\ell-1} - (\ell^2/4
+ q^2) z_k^2] \psi_{\ell-1} = 0$, and so one can replace
$(D^-_\ell D^+_{\ell-1})$ by $(\ell^2/4 + q^2)z_k^2$. 
Carrying this out repeatedly yields eq.~\eqref{Pellexp}.}
A decaying solution that goes as $1/z^{\ell+1}$  at large $z$ can be shown to give a non-vanishing $P_\ell$ as well.
It is worth noting that $P_\ell$, up to an $\ell$-dependent factor, is
the dissipative response identified by
\cite{LeTiec:2020spy,LeTiec:2020bos} (see also \cite{Wong:2019yoc} on
the case of the scalar).
This was worked out by analytically continuing $\phi_\ell$ ({\it not}
$\psi_\ell$) to non-integer $\ell$ and reading off the coefficient of
the resulting $1/r^{\ell+1}$ tail. 
It is interesting we obtain essentially the same quantity by computing
the conserved charge,
without the analytic continuation procedure.
As emphasized by
\cite{Chia:2020yla,Goldberger:2020fot,Charalambous:2021mea},
the dissipative response is distinct from the static, conservative
response. The latter is the Love number, the focus of this paper---it is defined by
the coefficient of the tidal response operator $\sim (\partial^\ell
\phi)^2$ in the point-particle
action~\cite{Goldberger:2005cd,Porto:2007qi,Kol:2011vg,
Hui:2020xxx,Goldberger:2020fot,Charalambous:2021mea}.
In this paper, we rely on these matching computations to relate the Love number to
the coefficient of the $1/r^{\ell+1}$ tail of $\psi_\ell$.\footnote{The connection between the conserved charge~\eqref{Pellexp} and the dissipative response suggests an alternative matching procedure that would proceed by matching these charges in the EFT.}

Of the two horizon asymptotics for $\psi_\ell$, the constant gives a non-vanishing $P_\ell$ while $\E^{-2iq \ln[(z-z_k)/z_k]}$
yields $P_\ell = 0$. Thus, charge conservation does not by itself tell us whether the decaying solution contains a
$\E^{-2iq \ln[(z-z_k)/z_k]}$ contribution when it is extended to the horizon (though it does tell us it must contain the constant contribution).
In order to figure out how the solution that decays at infinity behaves near the horizon, what we need is the Wronskian:
\be
W[\psi_\ell^{a} , \psi_\ell^{b}] = \E^{2iq \ln\left[{z-z_k\over z}\right]}
\big(\psi_\ell^{a} \Delta \partial_z \psi_\ell^{b} - 
\psi_\ell^{b} \Delta\partial_z \psi_\ell^{a}\big) \, ,
\ee
where $\psi_\ell^{a}$ and $\psi_\ell^{b}$ are any two
level-$\ell$ solutions. It can be shown $\partial_z W = 0$
(this is easiest to see by relating back to $\phi_\ell$ and inspecting eq.~\eqref{eomKerr}). 
We can choose $\psi_\ell^{a}$ to be the regular, growing solution, and
$\psi_\ell^{b}$ to be the decaying solution.\footnote{\label{decayingsoln}We do not need an explicit construction of the decaying
  solution for our argument, but here is how one could go about
  constructing one. The growing solution is the one obtained by raising
from $\psi_0 =\text{constant}$: $\psi_\ell = D^+_{\ell-1} \cdots D^+_0
\psi_0 \sim 1 + ... + z^\ell$. An independent solution that is simple
to construct is $\E^{-2iq {\,\rm ln\,}[(z-z_k)/z]}$ times the complex
  conjugate of the same $\psi_\ell$. A suitable linear combination of
  the two would give a $\psi_\ell$ that is purely decaying, going as
$1/z^{\ell+1}$ at large $z$. For instance, for $\ell=0$, the decaying
solution is proportional to $\E^{-2iq {\,\rm ln\,}[(z-z_k)/z]} - 1$. 
The higher $\ell$ decaying solution can be constructed by climbing the
ladder. (See eq.~\eqref{psi1psi2APP} in Appendix~\ref{hyperg} for the expressions of the independent solutions in terms of hypergeometric functions.)
}
Evaluating $W$ at large $z$ tells us $W$ is proportional
to the normalization of the $1/z^{\ell+1}$ tail.
The decaying solution, when extended to the horizon,
in general can have both the constant and the $\E^{-2iq \ln[(z-z_k)/z_k]}$ contributions, but the constant cannot contribute to $W$ (for $\psi_\ell^a$ also goes to a constant at the horizon)
while $\E^{-2iq \ln[(z-z_k)/z_k]}$ contributes to a non-vanishing $W$. By the constancy of $W$, we infer that
the decaying solution must be irregular at the horizon. 

Fact (1)---that the purely decaying solution is irregular at the
horizon---and fact (2)---that the regular solution is a finite
polynomial $\sim 1 + \cdots + z^\ell$---are thus established, leading to
the no-hair property and vanishing Love number for a linear, massless scalar around a Kerr
black hole. Furthermore, the same symmetry understanding as in the
Schwarzschild case applies here: the decaying solution spontaneously
breaks the horizontal ladder symmetry, i.e., $\delta \psi_\ell
\equiv Q_\ell \psi_\ell \ne 0$, while the growing (regular) solution
respects it. Next, we extend the argument to vector and tensor perturbations.

\newpage
\section{Ladder in Spin: From Scalar to Vector and Tensor}
\label{ladderSpin}

Finally, we want to treat the most general case of interest, a field of arbitrary (integer) spin $|s| \leq 2$ in a Kerr background. Such perturbations are described by the Teukolsky equation, which in
the static limit in BL coordinates is~\cite{Teukolsky:1972my,Teukolsky:1973ha}:
\be
\partial_r \left(\Delta \partial_r \phi_{\ell }^{(s)}\right) 
+ s(2r - r_s) \partial_r \phi_{\ell }^{(s)}  + \left( {a^2 m^2 + is(2r - r_s) am \over \Delta} -
   (\ell-s)(\ell+s+1) \right) \phi_{\ell }^{(s)} = 0 \,,
   \label{eq:genteukolskyeq}
\ee
where a (spin-weighted) spherical harmonic transform is assumed, and
the $m$ dependence of the Newman--Penrose variable $\phi_{\ell }^{(s)}$ is suppressed. 
Making the  substitution
\be
\phi_\ell^{(s)} \equiv \Delta^{-s} \E^{i q \ln [(z-z_k)/z]} \psi_\ell^{(s)} \, .
\ee
in analogy with eq.~\eqref{phipsi}, we have:\footnote{Note that here we have not multiplied by an additional factor of $-\Delta$ in order to define $\tilde H$. This is because the spin ladder structure is most simply phrased in terms of $\tilde H$. In Appendix~\ref{ladderTeukolsky} we describe the $\ell$ ladder structure, which uses $H$ defined analogously to the previous cases.}
\be
\label{tildeH}
 \tilde H_{\ell s} \psi_{\ell }^{(s)} = 0 \,, \quad{\rm where} \quad \tilde H_{\ell s} \equiv
\partial_z (\Delta \partial_z) 
+ \left[(z_k - 2z)s + 2 i q z_k\right] \partial_z 
- (\ell + s)(\ell - s + 1) \, . 
%
\ee
The large $z$ asymptotics are: $\psi_{\ell }^{(s)} \sim z^{\ell +s}$ and
$1/z^{\ell - s + 1}$. The near-horizon asymptotics are:
$\psi_{\ell }^{(s)} \sim$ constant and $\E^{(s-2iq ) \ln[(z-z_k)/z_k]}$. 
A ladder structure raising and lowering $\ell$ can be found (see
Appendix \ref{ladderTeukolsky}), but there is another ladder more
useful to us. We define the operators 
\be
E^- \equiv \partial_z \, , \qquad\qquad E^+_s \equiv \Delta\partial_z + s(z_k - 2 z) + 2iq z_k \, .
\label{eq:spinraiselowerops}
\ee
These are raising and lowering operators in the spin, $s$:
\be
E^- E^+_s - E^+_{s-1} E^- = -2s \, , 
\ee
\be
\tilde H_{\ell s+1} E^+_s = E^+_s \tilde H_{\ell s}\, , \qquad\qquad
\tilde H_{\ell s-1} E^- = E^- \tilde H_{\ell s} \, .
\ee
This implies that
given a solution $\psi_{\ell }^{(s)}$, one can generate $\psi_{\ell
}^{(s+1)} = E^+_s \psi_{\ell }^{(s)}$ and $\psi_{\ell }^{(s-1)} = E^-
\psi_{\ell }^{(s)}$.
We can use this spin ladder structure to construct conserved charges at each $\ell$ and $s$ 
from the known one at $\ell=0$, $s=0$ (eq.~\eqref{Q0Kerr}):
\be
\label{Q00}
\partial_z P_{0}^{(0)} = 0 \,, \quad{\rm where} \quad 
P_{0}^{(0)} \equiv (\Delta \partial_z + 2iq z_k) \psi_{0}^{(0)} \, .
\ee
For instance, the conserved charge for a general 
$\ell$ and $s=2$ solution $\psi_{\ell }^{(2)}$ is
\be
P_{\ell }^{(2)} \equiv (\Delta \partial_z + 2iq z_k) D^-_{1} \cdots D^-_{\ell} E^- E^- \psi_{\ell }^{(2)} \, ,
\ee
where the $D^-$'s are $\ell$-lowering operators for the scalar~\eqref{DKerr}. The two $E^-$ operators serve to lower the spin from $2$ to $0$, and the $D^-$ operators then lower $\ell$ to $0$.
The resulting $\psi_{0}^{(0)}$ can be plugged into
eq.~\eqref{Q00} to show $\partial_z P_{\ell }^{(2)} = 0$.

The growing branch of solutions for $s=2$ can be obtained by, once again,
starting from the scalar $\psi_{0}^{(0)} = {\rm constant}$ solution. Then
$\psi_{\ell }^{(2)} = E^+_1 E^+_0 D^+_{\ell-1} \cdots D^+_0 \psi_{0}^{(0)}$ 
is manifestly a finite polynomial of the form
$1 + z + \cdots + z^{\ell+2}$, judging from the form of $E^+$ and
$D^+$.\footnote{Note that $\psi_\ell^{(2)} \sim z^{\ell+2}$ at large $z$ means 
$\phi_\ell^{(2)} \sim z^{\ell-2}$ far away, as is appropriate for something
built out of second gradients of the metric.}
This implies that the purely growing at infinity solutions are regular at the horizon.

The decaying branch of solutions for $s=2$ can be obtained from
$E^+_1 E^+_0 \psi_{\ell }^{(0)}$ where $\psi_{\ell }^{(0)}$ is the level $\ell$ scalar decaying
solution (see footnote \ref{decayingsoln}). Recall that
$\psi_{\ell }^{(0)}$ scales as $1/z^{\ell +1}$ at large $z$ and contains a 
$\E^{-2iq \ln[(z-z_k)/z_k]}$ contribution as $z \rightarrow z_k$. 
Therefore, $E^+_1 E^+_0 \psi_{\ell }^{(0)}$ goes as 
$1/z^{\ell - 1}$ at large $z$ and contains $\E^{(-2iq+2)
  \ln[(z-z_k)/z_k]}$ close to the horizon. This means that the
decaying branch is irregular at the horizon in the same sense as
before.

We have thus established facts (1) and (2) for tensor perturbations
$s=2$, from which the no-hair property and vanishing  Love numbers follow.\footnote{To go from the Teukolsky solution $\psi_{\ell}^{(2)}$ to the Love numbers (really elements of the Love tensor), a matching computation is needed to connect the fall-offs of
  $\psi_{\ell}^{(2)}$ to the coefficients of the tidal response operators
  $\sim (\partial^{\ell-2} C)^2$ in the
  point-particle action, with $C$ the electric/magnetic part of
the Weyl tensor. See~\cite{Charalambous:2021mea,Goldberger:2020fot}. 
Similar comments as in footnote \ref{matching} apply here.
}
The same arguments translate straightforwardly to vector perturbations
$s=1$. It is important to emphasize the spin-weighted spherical
harmonics restrict us to $\ell \ge |s|$. Thus, the absence of hair and
Love numbers holds for $\ell \ge 2$  tensor perturbations, and $\ell \ge 1$
vector ones. 

The symmetry story works as before (recognizing that 
with the Teukolsky equation, we can only work with the equation of
motion as opposed to an action): the growing branch respects the
horizontal ladder symmetry while
the decaying branch spontaneously breaks it. Details are
given in Appendix \ref{ladderTeukolsky}.

It is worth mentioning that the spin raising and lowering operators~\eqref{eq:spinraiselowerops} are related to what are known
as Teukolsky--Starobinsky identities
\cite{Press:1973zz,Starobinskil:1974nkd}, which relate solutions to the Teukolsky equation with spin index $-s$ to those with $+s$.
In Chandrasekhar's notation
\cite{Chandrasekhar:1985kt}, the identities are:
$\phi^{(-1)} = \Delta {\cal D}_0^\dagger {\cal D}_0^\dagger \Delta
\phi^{(1)}$ and 
$\phi^{(1)} = {\cal D}_0 {\cal D}_0 \phi^{(-1)}$ for spin 1, and
$\phi^{(-2)} = \Delta^2 {\cal D}_0^\dagger {\cal D}_0^\dagger {\cal D}_0^\dagger {\cal D}_0^\dagger \Delta^2
\phi^{(2)}$ and 
$\phi^{(2)} = {\cal D}_0 {\cal D}_0 {\cal D}_0 {\cal D}_0
\phi^{(-2)}$ for spin 2, where ${\cal D}_0 \equiv \partial_r +
i[am -\omega (r^2
+ a^2)]/\Delta$, with $\omega$ being the frequency (i.e., $\phi \propto
\E^{-i\omega t}$). Here, we have kept the spin $s$  index explicit but suppressed
the $\ell$ (and $m$) dependence.
In the $\omega = 0$ limit, these are equivalent to, in our notation:
$\psi^{(-1)} = E^- E^- \psi^{(1)}$, $\psi^{(1)} =
E^+_0 E^+_{-1} \psi^{(-1)}$ for spin 1, and $\psi^{(-2)} = E^- E^- E^- E^-\psi^{(2)}$, $\psi^{(2)} =
E^+_1 E^+_0 E^+_{-1} E^+_{-2} \psi^{(-2)}$ for spin 2.
The new twist we are adding is this: in the static limit, we can
truncate these operations, enabling us to
increment $s$ by unity and still obtain a solution to the static Teukolsky equation, e.g.,
$\psi^{(0)} = E^- \psi^{(1)}$, $\psi^{(2)} = E^+_{1} \psi^{(1)}$, etc.
This is not possible for time-dependent solutions---with $\omega \ne 0$, only the
Teukolsky--Starobinsky combination of operations is useful, which connects $+s$ and
$-s$.

Here we have focused on the Teukolsky equation to see the symmetries underlying the computation of black hole Love numbers, but it is reasonable to ask whether similar structures can be seen at the level of metric perturbations (as opposed to Weyl scalars).
Around a Schwarzschild background this is indeed possible, as we discuss explicitly in Appendix~\ref{ladderReggeWheeler}.

\newpage
\section{Symmetries in the Infrared}
\label{IRsymmetries}

Our discussion so far has revolved around symmetries present in the
microphysical computation of static responses of black holes. However,
to a distant observer, such microscopic details of the black hole
geometry are invisible. Instead, they will infer the black
hole to be effectively a point particle, whose properties they can
probe for example by applying an external electric field, or
gravitational tidal field, and measuring the response. In this
effective description, the properties of the black hole are
encoded in an infinite number of coefficients appearing in the point-particle action; the values of these coefficients are determined by the requirement that the measured
responses are correctly reproduced. From this effective field theory (EFT) point of view, the vanishing of black hole static response coefficients is somewhat puzzling, because it would seem to be a fine-tuning~\cite{Rothstein:2014sra,Porto:2016zng,Penna:2018gfx}. It is natural to ask whether the ultraviolet symmetries we have identified can provide a symmetry explanation in the infrared, and in this section we  explore this question.

For concreteness, we will specialize to the simplest setting: the
worldline EFT of a massive (non-spinning) particle coupled to a scalar
field, which is the relevant effective theory describing, for example, the response of a Schwarzschild black hole to an externally applied scalar field profile.
The effective theory (up to quadratic order in $\phi$)
governing static field configurations takes the form of a bulk action plus
the point-particle worldline action:\footnote{{Strictly speaking the EFT also includes interactions of both the worldline and the bulk scalar with the bulk graviton. These couplings are not necessary to do the leading order matching, but are required to reproduce the full scalar field profile at subleading order in $1/r$. We comment further on the interplay between these graviton couplings and the symmetries that we have identified below.}}
\be
S = -\frac{1}{2}\int \rd^4 x \,(\partial\phi)^2+ \int \rd \tau \gamma
\bigg[\frac{1}{2}\gamma^{-2} \dot x^\nu\dot x_\nu - \frac{\mu^2}{2} -  g \phi\\
+\sum_{\ell=0}^{\infty} \frac{\lambda_\ell }{2\ell!} \left(  \partial_{(a_1}\cdots \partial_{a_\ell)_T} \phi \right)^2\bigg]\,,
\label{eq:pointscalaraction}
\ee
where $x^\nu$ denotes the spacetime position of the point particle,
with $\mu$ its mass. The variable $\gamma$ is a worldline vielbein
enforcing reparametrization invariance of the effective action and
$(\cdots)_T$ denotes the symmetrized traceless part of the enclosed
indices. The indices $a_1 \cdots a_\ell$ are defined by projecting space-time indices onto the
particle rest frame---see e.g.,~\cite{Hui:2020xxx} for an explicit definition of such projection. 

It is simplest to think of the particle as being
at rest at the origin, in which case a term like $g\phi$ has the
effect of adding a source term to the (static) $\phi$ equation of motion, i.e., $\nabla^2 \phi = g \delta_D$ where $\delta_D$ is the 3D
Dirac delta function centered at the origin. As such, $g$
can be thought of as the monopole scalar charge. 
If one wishes to describe a non-spherically-symmetric object, the possibility of higher multipole hair 
could be included by
adding linear terms of the form $\partial^\ell \phi$.
The couplings $\lambda_\ell$ are the response coefficients that encode the
particle's response to a static external field with multipole moment
$\ell$. Their presence means the static $\phi$
equation of motion takes the schematic form $\nabla^2 \phi \sim
\lambda_\ell \, \partial^\ell (\partial^\ell \phi \, \delta_D)$. Imagine the external
tidal field as the input $\phi$ on the right hand side; solving this
equation then gives the linear response, that is, an $r^\ell$ tidal
field sources an $r^{-(\ell+1)}$ response. Note that in
eq.~\eqref{eq:pointscalaraction} the metric is treated as flat, as is
appropriate at a large distance $r$. Beyond the lowest order, one would also need
to take into account couplings with metric perturbations, sourced by
the particle itself, whose effect can be included order by order
in $r_s/r$. 

A computation in the ultraviolet, such as those discussed in the previous sections,
can be compared against the infrared predictions to fix
the EFT coefficients $g$ and $\lambda_\ell$.  Indeed, we conclude that they
would vanish if the point particle were a black hole. But here, we wish to
ask a different question: what symmetries, visible in the infrared, might tell us that the
hair or tidal response coefficients should vanish? 

A possible candidate is the large distance limit of the ultraviolet
symmetries we have worked out---the ladder
symmetries. 
As we show in Appendix \ref{geometry},
the geometric perspective, in which a suitably rescaled and 
reduced Schwarzschild black hole is a Euclidean anti-de Sitter space in disguise, offers 
a particularly simple way to think about the whole ladder structure. 
The associated isometries
have a well defined flat space (i.e. infrared) limit, giving the symmetry
transformation in~\eqref{deltaphflat}:
\be
\delta\phi = r^2 \cos\theta \partial_r \phi
+ r  \partial_\theta (\sin\theta \phi) \, ,
\label{eq:flatsymm}
\ee
after scaling out an irrelevant overall factor.
The angular structure of this transformation mixes different $\ell$
values in spherical harmonic space, leading to the (large $r$ limit
of) $D_\ell^\pm$  operators defined in 
\eqref{eq:schwarzschildscalarladders}. The
transformation~\eqref{eq:flatsymm} may look unfamiliar, but it is
actually nothing more than a special conformal transformation, written
in spherical coordinates.\footnote{This can be made manifest by transforming to Cartesian coordinates and writing the transformation~\eqref{eq:flatsymm} in a rotationally covariant way, where it takes the more familiar form:
\be
\delta\phi = c_i \big( x^i -\vec{x}^2\partial^i+ 2 x^i \vec{x}\cdot \vec{\partial}\, \big)\phi \, ,
\label{eq:flatsymmcarts}
\ee
which is a special conformal transformation for a field with weight $1/2$. This equation reproduces~\eqref{eq:flatsymm} for $c_i=(0,0,1)$. Since we are
assuming that the theory is also rotationally invariant, the
symmetry~\eqref{eq:flatsymmcarts} with arbitrary $c_i$ is also
present.}

We now want to understand how the symmetry~\eqref{eq:flatsymm}, or equivalently~\eqref{eq:flatsymmcarts}, constrains the effective action~\eqref{eq:pointscalaraction}.
The bulk $(\partial\phi)^2$ term is clearly invariant in the static limit, but none of the worldline-localized couplings are invariant. This is more subtle than it might appear at first---since the transformation~\eqref{eq:flatsymmcarts} involves factors of $\vec{x}$, one may be tempted to conclude that the variation of an operator localized on the worldline vanishes when computed at the location of the point particle, $\vec{x}=0$. However, this makes a tacit assumption about the behavior of the field at $\vec x = 0$. This issue can be settled by showing that the would-be Noether currents associated with the symmetry transformation~\eqref{eq:flatsymm} in $\ell$ space are not conserved on-shell when any of the worldline couplings are present.\footnote{The non-invariance of the worldline couplings $\lambda_\ell$ also follows from the fact that they inevitably introduce a scale in the theory, thus breaking the conformal symmetry. This is because these terms capture finite-size effects~\cite{Ross:2012fc}.}
Thus,~\eqref{eq:flatsymm} is the
sought-after infrared symmetry which forbids Love number (and hair) couplings in the
point-particle effective theory. 

This appears to be a consistent story for black holes. How about stars, which in general can have a non-trivial hair and tidal
response? The symmetry~\eqref{eq:flatsymm} would appear to be broken
for them. How does that happen? To understand this, it is useful to recall the {\it
  horizontal} part of the ladder symmetries, effected by $Q_\ell$ 
defined in~\eqref{Q0def} and~\eqref{Qelldef}, for which a large $r$
limit also exists.\footnote{Given the relation of $D_\ell^\pm$ to conformal symmetry, it is tempting to speculate that the $Q_\ell$ symmetries for the free scalar in flat space
are related to known higher-spin symmetries, a subject for
further investigation.
}
Recall the purely growing mode solution $\phi_\ell^{({\rm g})}
\sim r^\ell + \cdots$ preserves the horizontal symmetry $Q_\ell
\phi_\ell^{({\rm g})} = 0$. On the other hand, any solution that has
an admixture of the decaying solution at infinity, $\phi_\ell^{({\rm
    d})} \sim 1/r^{\ell+1}+\cdots$, breaks the symmetry spontaneously,
in the sense that $Q_\ell \phi_\ell^{({\rm d})} \neq 0$. For stars, the
boundary conditions force us to include some contribution from this
decaying branch, leading to them having a nonzero tidal response. From
this viewpoint, we would expect the symmetry~\eqref{eq:flatsymm} to be
either realized nonlinearly in the EFT, or absent. In either case, it
is consistent that the EFT couplings that reproduce tidal responses
are not invariant under the linearly realized symmetry.

It is worth stressing that imposing the symmetry~\eqref{eq:flatsymm}
rules out  {\it all} multipolar hair and response couplings in one
fell swoop (as is appropriate for a 4D black hole in general relativity).
For a more general object, such as a star or a higher dimensional
black hole, it is conceivable the tidal response or
hair vanishes for some multipoles but not for others. 
In such a case, one would need a more flexible formulation of the 
ladder symmetries.  
 Decomposing $\phi$ in spherical harmonics,~\eqref{eq:flatsymm}
 implies a specific pattern of $\delta\phi_\ell$'s~\eqref{verticalSymmAll}. One could instead consider a modified
 version where only a particular neighboring pair, e.g.,
 $\delta\phi_\ell$ and $\delta\phi_{\ell-1}$, are non-vanishing~\eqref{pairDs}---this would be an example of the vertical symmetries
 restricted to a particular subset of $\ell$s. There are also the
 horizontal symmetries, one for each $\ell$: $\delta\phi_\ell =
 Q_\ell\phi_\ell$. Since these symmetries act on a single multipole,
 the structure allows---in principle---the breaking of the $Q_\ell$
 symmetries for some $\ell$s but not others. Exactly how this may manifest in the worldline EFT is something that we plan to explore in the future.

In this section, we have focused on the simplest situation where the hair and
the tidal field is a scalar, indeed a free scalar. We expect the
symmetry~\eqref{eq:flatsymm} to be broken {by generic} interactions. For instance, in our infrared discussion so far, we have
ignored gravitational interactions by working in the {leading $r\to \infty$} limit. {These interactions are in fact necessary to match the subleading $r_s/r$ corrections to the scalar profile.} 
What is interesting, perhaps surprising from the infrared point of view, is that {(a subset of the)} metric
perturbations can be systematically included, in fact resummed into the
Schwarzschild background, around which the symmetry does not
get destroyed, but elevated to~\eqref{deltaphirs}:
\be
\label{deltaphiRs}
\delta\phi = \Delta  \cos\theta \partial_r \phi
+ \left( r - {r_s \over 2}\right) \partial_\theta (\sin \theta \phi) \, .
\ee
We have thus come full circle to the ultraviolet symmetries that got
us started in this EFT discussion. {Based on the computation in
  the ultraviolet, we know that \eqref{deltaphiRs} is a symmetry of
  the infrared EFT if the metric perturbations were set to equal the
  ones needed to reconstruct the Schwarzschild metric. This should
  work order by order in an expansion in powers of $r_s/r$, of which 
  \eqref{eq:flatsymm} is the lowest order symmetry transformation. 
  We do not know, however, what the symmetry of the full infrared EFT
  is, when the metric perturbations can freely fluctuate. It is
  conceivable one could find a deformation of \eqref{eq:flatsymm}
  order by order in $r_s/r$, possibly supplemented by a corresponding
  transformation of the metric perturbations. This is left for future study.}

A number of questions remain to be
addressed. First, it would be interesting to work out in detail how the flat limit of our symmetries prevents black holes from having non-zero spin-1 and spin-2 response. In a separate direction, the case of a scalar around a rotating black hole is harder
to interpret from a geometric perspective (see Appendix
\ref{geometry}), but it remains true that the ladder symmetries 
have a well defined large distance limit.
The same can be said for vector or tensor perturbations around a Kerr
background. We expect that the ultraviolet symmetries we have identified will have similar consequences in the infrared, although the symmetries in question could only be guessed at
by inspecting the Teukolsky equation. A formulation in terms of the original metric perturbations rather than Newman--Penrose variables would be
desirable in order to draw a connection with the low-energy EFT of a
spinning point-particle~\cite{Porto:2005ac,Delacretaz:2014oxa}. 

Lastly, it is worth stressing again that existing no-hair theorems, which allow for e.g.,~scalar
self-interactions \cite{Bekenstein:1972ny,
  Bekenstein:1971hc,Bekenstein:1995un,Hui:2012qt}, 
are more powerful and generic than the version we focus on here.
Could our symmetry understanding
be adapted or generalized to those cases?
Or could the generic nature of black hole no-hair theorems be pointing us to
a completely different sort of explanation? One reason to search
  for a symmetry explanation is that in the worldline EFT of a
  spinning particle, a Kerr black hole will correspond to a specific
  pattern of multipolar couplings to
  gravity~\cite{Porto:2006bt,Levi:2015msa,Guevara:2020xjx}. Since
  black holes are such special objects, it is reasonable to ask if there
  is a symmetry that acts in the EFT that will guarantee these precise
  relationships between Wilson coefficients. On the other hand,
  in this case we know that the no-hair property of black holes is
  in a sense generic once one imposes regularity at the horizon as a
  boundary condition.

\section{Discussion}
\label{discuss}

We have seen that static linearized perturbations of massless fields in black hole backgrounds have a hidden structure of ladder-like symmetries and conserved quantities. These conserved quantities allow us to relate the behaviors of solutions near infinity to their behaviors near the black hole horizon {\it without} explicitly solving the relevant differential equations everywhere in between. This has useful implications for understanding both black hole no-hair theorems and the vanishing of black hole Love numbers from a symmetry perspective.

The question of whether or not a black hole can have hair is one that is naturally posed from very far away. We define hair to be the coefficient of the $1/r^{\ell+1}$ component of a static solution and the question of whether a black hole can have static hair is equivalent to asking whether or not a solution that has only this {asymptotic} behavior exists in a black hole background. We have seen that the conservation of charges associated to what we have called the horizontal ladder symmetries demands that this solution be singular at the horizon of a black hole---indicating that it cannot have {linear} hair. The nice feature of this argument is that it is based on symmetry, but it is noticeably less general than other no-hair theorems, which apply nearly irrespective of interactions, and essentially rely only on the existence of a horizon~\cite{Bekenstein:1972ny,Bekenstein:1971hc,Bekenstein:1995un,Hui:2012qt}. {In fact, the no-hair theorems are in a sense a confirmation that our generic expectation happens---that is, the solution with only a $1/r^{\ell+1}$ fall-off near infinity has an admixture of both fall-offs at the horizon, implying that it is singular there. In light of this, perhaps a symmetry explanation is unnecessary from the ultraviolet perspective. From the EFT point of view however, the no hair theorems imply a particular pattern of couplings, and one might wonder whether these can be explained by some symmetry.}


In contrast to the question of black hole hair, the question of how a
black hole responds to a static external field is naturally posed 
near the black hole horizon. In the vicinity of a black hole, we can
deduce that the two linearly independent solutions to the static
equations of motion of a field are such that one of them is regular at the horizon
and the other diverges. The generic expectation is that, if we were to
take the (unique) regular solution and expand it as $r\to\infty$,
there would be a linear combination of the two linearly independent
fall-offs at infinity (one growing and one decaying). We call the
ratio of the decaying solution to the growing solution the static
response, or Love number, and it tells us how the black hole responds
if we apply an external growing mode background subject to the
constraint that the field profile be regular at the horizon. We have
seen that for black holes, symmetry forces the solution that is
regular at the horizon to {\it only} have the growing fall-off at
infinity, meaning that black hole static response coefficients
vanish. For an ordinary object, we impose different boundary
conditions at the surface of the object, leading to a component of the
subleading, decaying fall-off at infinity, which spontaneously breaks the symmetries. 

Identifying these symmetries clarifies some aspects of the effective point particle description of black holes. In the effective theory, the vanishing of hair or static responses correspond to apparently fine-tuned points in parameter space. However, we have seen that (at least in the scalar case) the locus in parameter space where hair and response coefficients vanish is an enhanced symmetry point. Of course, this does not by itself guarantee that there should be a microphysical theory that actually gives rise to this effective theory. However, since we originally identified these symmetries in the ultraviolet, the microphysical model consisting of linearized fields in a black hole background leads precisely to this low-energy description.

The symmetry viewpoint that we are advocating proves to be useful beyond pure general relativity. Once we understand in the microscopic theory what symmetry is responsible for either the vanishing of Love numbers or for hair not being present, we can ask if additional interactions preserve or break these symmetries and infer their properties, without having to explicitly solve the equations. As an example, it is well-known that adding a coupling of a scalar field to the Gauss--Bonnet curvature invariant of the form $\phi R_{\rm GB}^2$ can lead to black hole hair. We can understand this from the symmetry viewpoint: to leading order (i.e., neglecting back-reaction), this coupling breaks the $\ell = 0$ horizontal ladder symmetry, and so now we are allowed to include operators in the effective theory that give rise to a black hole hair. (See Appendix~\ref{app:GB} for more details.) We expect that it will be similarly  possible to understand some broad features of other proposals for gravitational physics~\cite{Cardoso:2017cfl} from this perspective.

We have focused on the setting of exactly static perturbations in four spacetime dimensions for simplicity, but it is possible to relax these assumptions. In particular, the extension to finite (but small) frequencies is straightforward. The extension to other dimensions is similarly easy: as we have mentioned, the only essential difference is that the quantum number $\ell$ should be replaced with $\hat\ell = \ell/(D-3)$. The ladder structures will still be present, but only integer values of $\hat\ell$ will be connected via the ladder to the $\ell = 0$ equation of motion, which is necessary to be able to write down conserved charges and infer that Love numbers vanish. This is satisfying because it is known that Love numbers are actually nonzero for certain values of $\hat \ell$ in general dimension, but they do indeed vanish for integer $\hat \ell$ in all dimensions~\cite{Kol:2011vg,Hui:2020xxx}, completely consistent with the expectation coming from the ladder symmetries.

It is important to stress that the symmetries we have identified apply only to linearized fields in the background of a black hole, and in that regard they are deeply mysterious---why should they be present at all? It is tantalizing to speculate that they are the manifestation of some structure present in full general relativity, specialized to this simple setting. While we do not know what such a deeper explanation is, a possible hint in this direction is provided by the geometrization of the symmetries. As we discuss in Appendix~\ref{geometry}, the ladder symmetries that we have used to understand the properties of static solutions admit a geometric description, where the linearized fields propagate in an effective AdS background---with the isometries of this space leading to the ladder structure. A related promising avenue is to further investigate the algebraic structure underlying both the vertical and horizontal ladder symmetries, which is still somewhat mysterious.
We expect that understanding these geometric and algebraic structures more fully will lead to additional insights, hopefully into the nonlinear regime.

\vspace{-.3cm}
\paragraph{Acknowledgements}
We thank Marc Casals, Misha Ivanov, and Alberto
Nicolis for useful comments and discussions, and Justin Ripley for
pointing out the Teukolsky--Starobinsky identities. {We are especially grateful to Sergei Dubovsky for extensive discussions about the implications of our results for black hole solutions with scalar hair.} RP would like to acknowledge the hospitality of the Abdus Salam International Centre for Theoretical Physics (ICTP) during the final stage of this project. LH is supported by the DOE DE-SC0011941 and a Simons Fellowship in Theoretical Physics. The work of AJ is part of the Delta-ITP consortium. The work of RP is supported in part by the National Science Foundation under Grant No. PHY-1915611. LS is supported by Simons Foundation Award No. 555117. ARS is supported by DOE HEP grants DOE DE-FG02-04ER41338 and FG02-06ER41449 and by the McWilliams Center for Cosmology, Carnegie Mellon University.

\vspace{0.5cm}

\appendix

\newpage
\section{Geometric Origin of the Ladder Symmetries}
\label{geometry}

Some of the mystery surrounding the ladder symmetries introduced in the main text can be eliminated by viewing them as arising geometrically. Here we show how this works for a scalar field in a Schwarzschild background, showing that in the static limit it can be recast as propagating in an effective Euclidean anti-de Sitter space. The generalization  to the other cases will be discussed elsewhere.

Consider a static scalar $\phi$ in a Schwarzschild background.\footnote{For simplicity we restrict to completely static field configurations in this appendix. However, all of our manipulations can be extended to finite frequencies whose wavelength is large compared to the other scales in the problem (which is the near-horizon limit of~\cite{Bertini:2011ga,Charalambous:2021kcz}).}
Its action takes the form:
\be
\label{phiaction}
S = {1\over 2} \int \rd\theta \, \rd\varphi \, \rd r \sqrt{g} \, \phi 
  \square \phi \, ,
\ee
where the 3D metric, $g {}_{ij}$, is given by:
\be
\label{gS3D}
\rd s^2 = \rd r^2 + \Delta \left(\rd \theta^2 + \sin^2\theta \, \rd \varphi^2\right) \,
  \, .
\ee
Let us rescale the metric and the scalar by the following Weyl transformation
\be
\tl g_{ij} = \Omega^2 g_{ij} \, , \qquad \tl\phi =
  \Omega^{-{d-2 \over 2}} \phi \, ,
\ee
where $d = 3$ for our case of interest. The action transforms according to the following
identity:
\be
S = \frac{1}{2} \int \rd^d x \sqrt{g} \phi \square \phi = 
\frac{1}{2}\int \rd^d x \sqrt{\tl g} \left[ \tl \phi \tl\square \tl \phi- \Omega^{d-2 \over 2} {1\over \sqrt{\tl g}}
\partial_i \left(\sqrt{\tl g} \,\tl g^{ij} \partial_j \Omega^{-{d-2
  \over 2}}\right) \tl \phi {}^2 \right] \, .
\ee
Choosing $d=3$ and $\Omega = L^2/\Delta$, where
$L$ is an arbitrary length scale, it can be shown that the action takes the form
\be
S=\frac{1}{2}\int \rd^3 x \sqrt{g} \phi \square \phi = 
\frac{1}{2}\int \rd^3 x \sqrt{\tl g} \left( \tl \phi \tl\square \tl \phi
+ {r_s^2 \over 4 L^4} \tl\phi {}^2 \right) \, .
\label{eq:scalarinAdS}
\ee
This means that 
the rescaled scalar $\tl \phi = (\sqrt{\Delta} / L) \phi$
behaves as a massive scalar in the following background:
\be
\tl{\rd s}^2 =\left(\frac{L^2}{\Delta}\right)^2 \Big(\rd r^2 + \Delta \rd\theta^2 + \Delta \sin^2\theta \,
  \rd\varphi^2 \Big) \, .
  \label{eq:eads1}
\ee
Though it is not obvious, this is a constant curvature space, with Ricci scalar
\be
\tl R_{\rm EAdS} = -\frac{3r_s^2}{2L^4}\,.
\ee
Since the curvature is negative, it is hyperbolic space (Euclidean AdS${}_3$). It might seem somewhat peculiar that the scalar field, which is {\it not} conformally coupled,\footnote{One way to check this is to note that the scalar's mass in EAdS is $m^2 = - \frac{1}{6}\tl R_{\rm EAdS}$, which is the $d=4$ conformally coupled value, instead of the $d=3$ conformally coupled value $m^2 = - \frac{1}{8}\tl R_{\rm EAdS}$.}  can be mapped via a Weyl transformation to a scalar with a constant mass in EAdS. The underlying reason for this is that the action for a scalar in the full four-dimensional Schwarzschild spacetime is conformally coupled, as we will explain elsewhere.

Regardless of its origin, the fact that~\eqref{phiaction} is related by a Weyl transformation to a scalar theory in EAdS~\eqref{eq:scalarinAdS} implies that the isometries of EAdS (which act as conformal Killing vectors on the original metric) have a natural action on $\phi$: 
\be
\delta\phi = \xi^i\partial_i\phi+\frac{d-2}{2d}\nabla_i\xi^i\phi\,,
\label{eq:confkilling}
\ee
where $\xi^i$ are conformal Killing vectors of the metric $g_{ij}$ and true Killing vectors of the metric $\tl g_{ij}$.
Conceptually, the first part of this transformation comes from the ordinary coordinate transformation that is a conformal Killing isometry, while the second part comes from the compensating Weyl transformation needed to bring the metric back to its original form.

What are the isometries of $\tl g_{ij}$? 
They are easier to recognize
by defining a new radial coordinate:
\begin{align}
 r_* &\equiv {L^2 \over r_s} \ln {r-r_s \over r} \, , & 
\rd r_* &\equiv {L^2 \rd r \over \Delta}  \, , \nonumber \\
 r &= {r_s \over 1 - \E^{r_* r_s/L^2}} \, ,  &
\Delta &= {r_s^2 \over 4} \frac{1}{ \sinh^2 (r_* r_s/(2 L^2))} \, ,
\end{align}
such that the line element~\eqref{eq:eads1} is cast into the form of EAdS${}_3$ in global coordinates
\be
\tl{\rd s} {}^2 = \rd r_*^2 +  {4 L^4 \over r_s^2} \, 
\sinh^2 \left({r_* r_s \over 2L^2}\right)
  (\rd\theta^2 + \sin^2\theta \rd\varphi^2) .
\ee
The space has 6 Killing vectors: 3 rotations and 3 translations
(or ``boosts''). As an example, consider the translation that mixes
$r_*$ and $\theta$, i.e., the Killing vector (in $\tl
g_{ij}$ sense):
\be
J_{r_*\,\theta} = -{2L^2} \cos\theta \, \partial_{r_*} + 
r_s \coth\left( {r_* r_s \over 2 L^2} \right) \sin\theta 
\, \partial_\theta \, .
\ee
From this, we can extract the vector $\xi^i$. It is a conformal Killing vector of the original metric~\eqref{gS3D} and its action on the scalar is then given by~\eqref{eq:confkilling}
\be
\label{deltaphirs}
\delta\phi = -{2} \Delta \cos\theta \partial_r \phi
+ (r_s - 2r) \partial_\theta (\sin\theta \phi) \, .
\ee
Note that
the arbitrary scale $L$ disappears
from $\delta\phi$, as it should.
Decomposing $\phi$ into spherical harmonics, it can be shown that this symmetry acts on the various $\ell$ components as~\cite{Compton:2020cjx}
\be
\label{verticalSymmAll}
\begin{aligned}
 \delta \phi_\ell &= f(\ell-1) D^+_{\ell-1} \phi_{\ell-1} - f(\ell)
  D^-_{\ell+1} \phi_{\ell+1}  \, , \\
\delta \phi_{\ell-1} &= f(\ell-2) D^+_{\ell-2} \phi_{\ell-2} - f(\ell-1)
  D^-_{\ell} \phi_{\ell}   \, , \\
 f(\ell) &\equiv \left(\frac{4(\ell+1 - m) (\ell+1+m)}{ 3 +
   4\ell(\ell+2)} \right)^\frac{1}{2}\, ,
\end{aligned}
\ee
where the $D_\ell^\pm$ operators appearing are precisely the ones in~\eqref{eq:schwarzschildscalarladders}.
We highlight 
the pair $\delta\phi_\ell$ and $\delta\phi_{\ell-1}$ (among
the whole tower of $\delta\phi$'s) to
emphasize that the invariance of the total action comes
from the cancellation between the two $f(\ell-1)$ terms.
It is useful in fact to zero in on the pair
$\phi_\ell$ and $\phi_{\ell-1}$ (ignoring $\phi_{\ell+1}$ 
and $\phi_{\ell-2}$ contributions to $\delta\phi_\ell$ and
$\delta \phi_{\ell-1}$). 
In spherical harmonic space, the action is a sum over $\ell$ of the radial integral of
$\phi_\ell^* H_\ell \phi_\ell$, the variation of the $\ell^\mathrm{th}$ term is
\be
\delta(\phi_\ell^* H_\ell \phi_\ell) = f(\ell-1) \left[ \phi_\ell^* H_\ell D^+_{\ell-1} \phi_{\ell-1}
+ (D^+_{\ell-1} \phi_{\ell-1})^* H_\ell \phi_\ell\right] \, ,
\ee
and the variation of the $(\ell-1)^\mathrm{th}$ term, $\phi_{\ell-1}^* H_{\ell-1} \phi_{\ell-1}$ is
\be
\delta(\phi_{\ell-1}^* H_{\ell-1} \phi_{\ell-1}) = -f(\ell-1) \left[\phi_{\ell-1}^* H_{\ell-1} D^-_\ell \phi_\ell
+ (D^-_\ell \phi_\ell)^* H_{\ell-1}
\phi_{\ell-1} \right] \, .
\ee
One can see these terms cancel, upon recognizing that
$H_\ell D^+_{\ell-1} = D^+_{\ell-1} H_{\ell-1}$ and
$H_{\ell-1} D^-_{\ell} = D^-_{\ell} H_\ell$,
as well as
$D^-_{\ell} {}^* \phi_\ell^* = \phi_\ell^* D^+_{\ell-1}$ 
and $D^+_{\ell-1} {}^* \phi_{\ell-1}^* = \phi_{\ell-1}^* D^-_\ell$
from integration
by parts. The precise form of 
$f(\ell-1)$ does not matter, except 
one feature: that $f(\ell-1)$ implicitly depends on $m$
and vanishes when $m= \pm \ell$. This prevents the generation of terms
with $\ell < |m|$, and allows the whole tower to terminate at $\ell=0$.
The pairing up of $\phi_\ell$ and $\phi_{\ell-1}$ in this way points
to a supersymmetric structure, spelled out in Appendix \ref{SUSY}.
We refer to this symmetry that mixes levels as a {\it vertical} ladder symmetry.

The charge conservation we use in Section \ref{ladderSchw} arises from
the {\it horizontal} symmetry transformation $\delta \phi_\ell = Q_\ell \phi_\ell$, 
which does {\it not} mix different $\ell$'s.
It arises from the simple observation that $Q_0 \equiv
\Delta \partial_r$ is a symmetry at the $\ell=0$ level;
symmetries at higher $\ell$'s are built up from $Q_0$ by climbing 
the ladder.

It is worth noting that this whole
geometric perspective has a well-defined flat space
$r_s \rightarrow 0$ limit under which: $\Omega = L^2/r^2$, 
$r_* = -L^2/r$, 
$\tl{\rd s}{}^2 = \rd r_*^2 + r_*^2 
(\rd\theta^2 + \sin^2\theta \rd\varphi^2)$,
and $\delta\phi$ becomes\footnote{In fact, in this limit the
    symmetry~\eqref{deltaphflat} is nothing but a special conformal
    transformation~\eqref{eq:flatsymmcarts}. It would be very interesting to understand in more detail the relation between these flat space conformal symmetries and the conformal Killing symmetries in the Schwarzshild background.}
\be
\label{deltaphflat}
\delta\phi = - 2r^2 \cos\theta \partial_r \phi
- 2r \partial_\theta (\sin\theta \phi) \, ,
\ee
which is relevant for our discussion in Section~\ref{IRsymmetries}.

In the Kerr case, the static scalar 
effectively
lives in the metric $g^K_{ij}$ given by
\be
\rd s_K^2 = {\rho^2 - r r_s \over \Delta} 
\left( \rd r^2 + \Delta \rd\theta^2 + 
{\Delta^2 \sin^2\theta \over \rho^2 - r r_s} \rd\varphi^2 \right)
  \, .
\ee
There does not seem to be a rescaling which brings this to (E)AdS
form. However, the action is not much more complicated than
that for the Schwarzschild case:
\be
S_K = \frac{1}{2}\int\rd^3 x\sqrt{g^K} \phi \square_K \phi =
\frac{1}{2}\int\rd^3 x \left( \sqrt{g} \phi \square \phi \right) - {a^2
  \sin\theta \over \Delta} \phi \partial_\varphi^2\phi \, ,
\ee
where $g_{ij}$ and $\square$ on the right hand side refers to
the metric in eq.~\eqref{gS3D}. 
Suppose we apply the same rescaling as before:
$\tl g_{ij} = (L^4/\Delta^2) g_{ij}$ and $\tl \phi =
(\sqrt{\Delta}/L) \phi$, then
\be
S_K=\frac{1}{2}\int\rd^3 x\sqrt{g^K} \phi \square_K \phi = \frac{1}{2}\int\rd^3 x\sqrt{\tl g} 
\tl \phi \left( \tl\square  + {r_s^2\over 4 L^4} - {a^2 \over L^4} 
\partial_\varphi^2 \right)    \tl \phi \, .
\ee
Separating the $\partial_\varphi^2$ term from the other derivatives
might seem strange, but the point is that in spherical harmonic
space, $\partial_\varphi^2 = -m^2$, and so the Kerr geometry merely
modifies the mass term for $\tl \phi$. Because of this,
a scalar in the Kerr background has the same ladder
structure (with modified $D^\pm$'s) as a scalar in the 
Schwarzschild background.

The geometric view of the ladder symmetries can be extended beyond spin 0, as well as non-vanishing frequency. This will
be discussed in a separate paper.

\newpage
\section{P\"oschl--Teller Potential}
\label{app:PT}

We can understand the simplicity of the static scalar in Schwarzschild by relating it to the well-known class of P\"oschl--Teller potentials~\cite{PTpotential}. Starting from the equation of motion for the scalar in Schwarzschild in the form~\eqref{eq:schscalarEOM}
\be
 \Delta \big(\partial_r (\Delta \partial_r) - \ell (\ell + 1)\big)\phi_\ell = 0\,,
 \label{eq:appschS}
\ee
we make the change of variables $\rd z_\star = r_s\Delta^{-1} \rd r$ which implies
\be
z_\star = 2\arctanh\left(\frac{r_s}{r_s-2r}\right).
\ee
After this change of variables, the equation~\eqref{eq:appschS} takes the form of a Schr\"odinger equation with a potential
\be
\left(\partial_{z_\star}^2 -\frac{\ell(\ell+1) }{4}\csch^2\left(\frac{z_\star}{2}\right)\right)\phi_\ell = 0\,,
\ee
which is precisely of the P\"oschl--Teller form.\footnote{Note that often the P\"oschl--Teller potential is written in terms of $\sech$ rather than $\csch$, but the two forms can be transformed into each other~\cite{Anninos:2011af}. The $\csch$ structure appears here because we are choosing $z_\star$ to cover the region of $r$ that lies outside the black hole.} 
This class of potentials has a number of very interesting features. Famously, for $\ell \in {\mathbb Z}$, scattering in the P\"oschl--Teller potential is reflectionless---a wave sent in from one side of the potential barrier passes through the barrier with no reflected amplitude. The vanishing of a black hole's tidal response is essentially a zero-frequency version of this reflectionless property: the physically relevant solution has only a single fall-off in both asymptotic regions. Interestingly the P\"oschl--Teller potential also plays a prominent role in discussions of the transparency of odd-dimensional de Sitter spaces~\cite{Lagogiannis:2011st,Compton:2020cjx}, revealing a surprising mathematical connection between black hole physics and de Sitter space.\footnote{In fact, in Appendix~\ref{geometry} we saw that one can understand the black hole ladder symmetries geometrically in terms of a scalar field propagating in an effective Euclidean AdS geometry, which is simply related by analytic continuation to the de Sitter problem.}

The transformation to the $z_\star$ variable also makes manifest  some features of the solutions. By inspection it is clear the two $\ell=0$ solutions are $\phi_0\sim {\rm constant}$ and $\phi_0\sim z_\star$, valid everywhere. Transforming back to the $r$ coordinate, the $z_\star$ solution is the one that diverges at the horizon (which is mapped to $z_\star = -\infty$).

The P\"oschl--Teller potential has a well-studied connection to supersymmetric quantum mechanics (which we elaborate on in Appendix~\ref{SUSY}), where the ladder operators~\eqref{eq:schwarzschildscalarladders} are essentially the supercharges. The potential is in particular an example of a shape-invariant potential, since potentials with consecutive integer values of $\ell$ are superpartners~\cite{Cooper:1994eh}.

Here we have shown how the scalar in Schwarzschild can be cast in the P\"oschl--Teller form, but the more general Teukolsky equation~\eqref{eq:genteukolskyeq} can also be put in a generalized P\"oschl--Teller form. This class of potentials has an algebraic structure equivalent to the one that we have discussed~\cite{Barut_1987,Wu:1989qoa, 1990JMP....31..557W}.

\newpage
\section{Ladder in Regge--Wheeler and Zerilli}
\label{ladderReggeWheeler}

In the main text we argued for the vanishing of Love numbers using the symmetries of the Teukolsky equation. One might wonder, however, if it is possible to see the phenomenon directly at the level of metric perturbations. Here we show that in  Schwarzschild background a similar structure is present in the Regge--Wheeler and Zerilli equations.
We start by considering the Regge--Wheeler equation.
The action governing the linearized dynamics of odd-parity spin-2 perturbations around Schwarzschild black holes in general relativity is (see \cite{Hui:2020xxx} for a detailed derivation in general spacetime dimensions)
\be
S_{\text{RW}} = \sum_\ell \int\D t \D r_\star \left( \frac{1}{2}\dot{\Psi}^{2}_{\ell} - \frac{1}{2} \left( \frac{\partial\Psi_{\ell}}{\partial r_\star} \right)^2 - \frac{1}{2}V_{\text{RW}}(r) \Psi_{\ell}^2 \right) \, ,
\label{RWaction}
\ee
where $V_{\text{RW}}(r)$ is the Regge--Wheeler potential 
\be
V_{\text{RW}}(r) = f(r) \left( \frac{\ell(\ell+1)}{r^2} -  f'(r)\frac{3}{r} \right) \, ,
\qquad
f(r) \equiv 1- \frac{r_s}{r} \, , 
\ee
and where $r_\star$ is the standard tortoise coordinate, defined as $\D r_\star \equiv \D r/ f(r) = r^2 \D r / \Delta$. Note that $\Psi_\ell$ in~\eqref{RWaction} is not quite the usual Regge--Wheeler variable \cite{Regge:1957td}, which is, instead, up to a numerical factor, the time derivative of our $\Psi_\ell$ \cite{Hui:2020xxx}.
In~\eqref{RWaction} we have integrated over the angles to remove the dependence on the spherical harmonics. 
In the static limit, the Regge--Wheeler equation, obtained from the action~\eqref{RWaction}, is
\be
\frac{\D^2\Psi_\ell}{\D r_\star^2}  - V_{\text{RW}}(r) \Psi_\ell =0 \, .
\label{RWeqstaticlimit}
\ee
One can introduce the following operators
\begin{align}
D_\ell^+ & \equiv  -\Delta \partial_r +  \frac{\ell^2+3}{2 (\ell+1)}r_s-\ell r    \, ,
\label{defDplusRW}
\\
D_\ell^-  & \equiv  \Delta \partial_r + \frac{\ell^2 (r_s-2 r)-2 \ell (r-r_s)+4r_s}{2 \ell}   \, ,
\end{align}
which satisfy 
\begin{equation}
D_{\ell+1}^-D_\ell^+ - D_{\ell-1}^+D_\ell^- = (2\ell+1) r_s^2\left[\frac{1}{4} - \frac{4}{\ell^2(\ell+1)^2}   \right] .
\end{equation}
It can be shown that the equations of motion~\eqref{RWeqstaticlimit} can be equivalently written as $r^{-4}H_\ell \Psi_\ell =0$, where
\be
H_\ell  \equiv D_{\ell-1}^+D_\ell^- - \frac{(\ell^2-4)^2r_s^2}{4\ell^2}= D_{\ell+1}^-D_\ell^+ - \frac{(\ell^2+2\ell-3)^2r_s^2}{4(\ell+1)^2} \, .
\ee
The algebra of $D_\ell^\pm$ with $H_\ell$ is the same as~\eqref{HD}: $H_{\ell+1}D_\ell^+ =  D_\ell^+H_\ell $ and $ H_{\ell-1}D_{\ell}^-=D_{\ell}^-H_\ell$, which implies that $D_\ell^+$ and $D_\ell^-$ are raising and lowering operators in the sense that, if $\Psi_\ell$ solves $H_\ell \Psi_\ell=0$ for some $\ell$, then $H_{\ell+1}(D_\ell^+\Psi_\ell)=0$ and $H_{\ell-1}(D_\ell^-\Psi_\ell)=0$. One can then repeat the analysis of Section~\ref{ladderSchw}. In particular, by direct inspection of the equations of motion for modes with $\ell=2$, we can identify the quantity $P_2 \equiv r^4 \Delta \partial_r (\frac{\Psi_2}{r^3})$ which is conserved on-shell, i.e., $\partial_r P_2=0$ if $H_2\Psi_2=0$. From the conservation of $P_2$ at large distances, one can find the  large-$r$ behavior of the two independent  solutions  to the Regge--Wheeler equation~\eqref{RWeqstaticlimit}: one is $\Psi_2^{(1)}(r\rightarrow\infty) \sim r^3$ (which is actually a good solution everywhere and corresponds to $P_2 =  0$), while the other goes as $\Psi_2^{(2)}(r\rightarrow\infty) \sim r^{-2}$ (which corresponds to $P_2= \text{constant}\neq 0$). From the conservation of $P_2$, one then infers that only $\Psi_2^{(1)}$ is regular at the black hole horizon, while $\Psi_2^{(2)}$  has to diverge logarithmically as $\ln (1-r_s/r)$. One can then generate regular solutions for all $\ell$'s by repeated use of the raising operator $D_\ell^+$, $\Psi_\ell^{\text{reg}}= D_{\ell-1}^+\cdots D_2^+ \Psi_2^{(1)}$. As is manifest from the definition~\eqref{defDplusRW}, $\Psi_\ell^{\text{reg}}$ is a finite polynomial with only positive powers of $r$. Then, using the constraint equations to compute the  metric fluctuation $\delta g_{t\phi}$ from the solution $\Psi_\ell^{\text{reg}}$, or, alternatively, performing the matching with the point-particle EFT~\eqref{eq:pointscalaraction} \cite{Hui:2020xxx}, one concludes that  the odd-type Love numbers vanish for all $\ell$'s.

In analogy with the scalar case, conserved quantities at higher $\ell$ can be obtained by climbing the ladder:
\be
\partial_r P_\ell = 0 \,,\quad {\rm where} \quad P_\ell \equiv r^4\Delta\partial_r (r^{-3}D^-_1 D^-_2 \cdots D^-_{\ell} \Psi_\ell) \, .
\ee
We stress that $P_2$---which was inferred from the structure of the Regge--Wheeler equation for $\ell=2$---could have been equivalently derived from  the action~\eqref{RWaction}. Indeed, it can be shown that, in the static limit, where one neglects time derivatives acting on $\Psi$, the sector with $\ell=2$ in~\eqref{RWaction}   is  invariant, up to total derivative terms, under the  transformation $\delta\Psi_2 = Q_2 \Psi_2$, where $Q_2 \equiv r^4 \Delta \partial_r -3 r^3 \Delta $. Following the Noether procedure, one  finds the conserved current $J_2= -\frac{1}{2} r^8\Delta^2[\partial_r (\frac{\Psi_2}{r^3})]^2 = -\frac{1}{2}P_2^2 $. Thus, the conservation of $J_2$ ($\partial_{r_\star}J_2=0$) is equivalent to $\partial_rP_2=0$. The procedure can be generalized to all $\ell$'s. In particular, one can introduce recursively the quantities
\begin{equation}
Q_\ell \equiv D^+_{\ell-1}  \, Q_{\ell-1}  \, D^-_{\ell} \, ,
\label{QellRW}
\end{equation}
which can be shown to leave the action~\eqref{RWaction} invariant. The operators~\eqref{QellRW} should be understood as follows: starting from a solution $\Psi_\ell$ at level $\ell$, the chain of  lowering operators $D^-_{3} \cdots D^-_{\ell}$ generate a solution at level $\ell=2$; $Q_2$ is then used to produce another solution with $\ell=2$; finally, the raising operators $D^-_{\ell-1} \cdots D^-_{2}$ yield a level-$\ell$ solution.

The previous argument has shown that the vanishing of the black hole Love numbers of odd type can be traced back to the existence of symmetries  governing the static regime of the Regge--Wheeler equation.  Similar considerations can be in principle extended and applied also to the case of the Zerilli equation~\cite{Zerilli:1971wd} and the corresponding Love numbers of even type. The logic that we will follow here is, however, slightly different. It is in fact much more convenient to understand the  vanishing of the even-type Love numbers by combining the previous result with  another property  of black hole perturbations in general relativity, which avoids  going through the derivation  of  ladder operators for the Zerilli equation.
We shall use instead the existence of a duality symmetry relating perturbations of different parity. This duality, which is the basis of   isospectrality of massless spin-2 fields around Schwarzschild black holes in general relativity \cite{10.2307/78870},\footnote{For a generalization to the case of partially massless spin-2 fields on Schwarzschild-de Sitter spacetimes, see \cite{Rosen:2020crj}.} can be shown to be also responsible for the \textit{equality} of the even and odd Love numbers, at any given $\ell$ \cite{uspaper2}.
Combining the equality of the Love numbers, as a result of the duality, with the vanishing of the static response for the Regge--Wheeler equation, which we inferred  above from the ladder symmetries, one can then conclude immediately that the  Love numbers of even type in general relativity must also vanish.

We conclude by mentioning that similar considerations and results, following from the existence of ladder operators that raise and lower $\ell$, straightforwardly apply also to the case of spin-$1$ perturbations around Schwarzschild black holes \cite{Hui:2020xxx}.

\newpage
\section{Ladder in Teukolsky}
\label{ladderTeukolsky}

In the main text, we were able to discern the properties of solutions to the Teukolsky equation by exploiting a ladder structure in spin, which relates solutions of the static spinning Teukolsky equation to its scalar counterpart.
However, there is also a more direct route, which is to identify a ladder in $\ell$ structure in the Teukolsky equation at fixed $s$. In this Appendix we exhibit this structure directly.

In order to see this ladder structure, it is convenient to 
multiply eq.~\eqref{tildeH} by $-\Delta$, and
rewrite the equation of motion as:
\be
H_{\ell s} \psi_{\ell }^{(s)} = 0 \, , \quad {\rm with} \qquad H_{\ell s} \equiv -\Delta \tilde H_{\ell
  s} \, .
\ee
Define the $\ell$ raising and lowering operators:
\be
\begin{aligned}
D^+_{\ell s} &\equiv -\Delta \partial_z + {\ell - s + 1\over 2(\ell +1)}\big [(\ell+1)(z_k - 2 z) - 2 iq z_k\big] \, ,  \\
D^-_{\ell s} &\equiv \Delta \partial_z + {\ell +s \over 2\ell} \big[\ell(z_k - 2z) + 2iq z_k\big] \, .
\end{aligned}
\label{eq:teukolskyladderD}
\ee

These operators obey the algebraic relations:
\be
\begin{aligned}
 D^-_{\ell+1 s} D^+_{\ell s} - D^+_{\ell-1 s} D^-_{\ell s} &= {(2\ell+1)
  z_k^2 \over 4} + {q^2 s^2 z_k^2 (2\ell + 1) \over \ell^2 (\ell+1)^2}
  \, , \\
 H_{\ell s} &= D^-_{\ell + 1 s} D^+_{\ell s}  - (\ell+s+1)(\ell-s+1){((\ell +1)^2 +
   4q^2)z_k^2 \over 4(\ell + 1)^2} \, ,\\
 &= D^+_{\ell-1 s} D^-_{\ell s} - (\ell+s)(\ell-s){(\ell^2 +
   4q^2) z_k^2 \over 4\ell^2} \, ,
\end{aligned}
\ee
and as before $D_{\ell s}^\pm$ act as ladder operators on the Hamiltonian:
\be
 H_{\ell + 1 s} D^+_{\ell s} = D^+_{\ell s} H_{\ell s} \,, \qquad
H_{\ell-1 s} D^-_{\ell s} = D^-_{\ell s} H_{\ell s} \, .
\ee
The appearance of non-local looking terms, factors of $\ell$ in the
denominators of~\eqref{eq:teukolskyladderD}, should not surprise us. Recall that for $s=1,2$, the Newman--Penrose
variables are first and second derivatives of the gauge fields
respectively (thus the Teukolsky equation involves more than two
derivatives on the fundamental fields). 
Ignoring this subtlety, 
we can still think of the raising and lowering operators as
giving us vertical symmetries of the equation of motion:
i.e., given $\psi_{\ell }^{(s)}$ which solves $H_{\ell s} \psi_{\ell }^{(s)} = 0$, 
$D^+_{\ell s} \psi_{\ell }^{(s)}$ satisfies $H_{\ell+1 s} (D^+_{\ell s}
\psi_{\ell }^{(s)}) = 0$, and $D^-_{\ell s} \psi_{\ell }^{(s)}$ satisfies $H_{\ell-1 s} (D^-_{\ell s}
\psi_{\ell }^{(s)}) = 0$. 
The fundamental origin of these symmetries is a bit obscure in this case, because we do not have an action for the Teukolsky equation. Nevertheless, they provide a useful device for manipulating solutions to the equation.

Proceeding in the same formal manner, we can guess
the horizontal ladder symmetries.
It is simplest to start with a negative $s$, i.e., $s=-1$ or $s=-2$
($s=0$ reduces to the scalar case).
In particular, for $\ell = -s = |s|$, the constant term in $\tilde H_{\ell s}$
vanishes, in which case the equation of motion
\be
\beta^2 H_{|s|s} \psi_{|s|}^{(s)} = 0 \, , \qquad {\rm with}\qquad
\beta \equiv \Delta^{-s} \E^{2iq \ln [(z-z_k)/z]} \, ,
\ee
takes the simple form:
\be
- \beta \Delta \partial_z (\beta \Delta \partial_z) \psi_{|s|}^{(s)} = 0
  \, .
\ee
It is easy to see that a symmetry of this equation is
\be
\delta \psi_{|s|}^{(s)} \equiv Q_{|s|}^{(s)} \psi_{|s|}^{(s)}\,,\quad {\rm where}\quad Q_{|s|}^{(s)}
\equiv \beta \Delta \partial_z\,.
\ee
This transformation is a symmetry
in the sense that
$[\beta^2 H_{|s|s} , Q_{|s|}^{(s)}] = 0$, so that 
if $\psi_{|s|}^{(s)}$ is a solution, so is $Q_{|s|}^{(s)} \psi_{|s|}^{(s)}$. We can
bootstrap to higher $\ell > -s = \vert s\vert$ by:
\be
Q_{\ell }^{(s)} =  D^+_{\ell-1 s} \cdots      D^+_{\vert s\vert s} \, Q_{|s|}^{(s)}  \,
  D^-_{\vert s\vert +1 s} \cdots D^-_{\ell s}  \, .
\ee
Starting from a level $\ell$ solution $\psi_{\ell }^{(s)}$, the combination
$D^-_{\vert s\vert+1 s} \cdots D^-_{\ell s} \psi_{\ell }^{(s)}$ produces a solution at
level $\ell = -s = |s|$. We already know acting on it with $Q_{|s|}^{(s)}$
produces another level $\ell = -s = |s|$ solution. Finally raising the result back with
$D^+_{\ell-1 s} \cdots D^+_{\vert s\vert s}$ produces another level $\ell$ solution.
We thus have an equation-of-motion symmetry in the sense that 
if $\psi_{\ell }^{(s)}$ is a solution, so is
$Q_{\ell }^{(s)} \psi_{\ell }^{(s)}$.
Note, however, that $[\beta^2 H_{\ell s}, Q_{\ell }^{(s)}] \ne 0$. 

Considering
 $s=-2$ for instance, the growing solution for $\ell = 2$  is
$\psi_{2}^{(-2)} = E^- E^- D^+_{1} D^+_{0} \psi_{0}^{(0)}$
with $\psi_{0}^{(0)}= {\rm constant}$. 
(Here, the $D^\pm$'s are the raising and lowering operators for 
scalar.) It is not hard to show $\psi_{2}^{(-2)}$ is a
constant too, and thus $Q_{2}^{(-2)} \psi_{2}^{(-2)} = 0$. 
Conversely, the decaying solution for $\ell=2$ is $\psi_{2}^{(-2)} = E^- E^- D^+_{1} D^+_{0} \psi_{0}^{(0)}$ 
with $\psi_{0}^{(0)} = \E^{-2iq\ln[(z-z_k)/z]}-1$. It can be checked that
$Q_{2}^{(-2)} \psi_{2}^{(-2)} \ne 0$.
The obvious extensions of these statements to $\ell > 2$ can
be derived by climbing the ladder. 

What about $s=2$? The corresponding symmetry operator can be obtained
by climbing the spin ladder:
\begin{equation}
Q^{(2)}_\ell = E^+_1 E^+_0 E^+_{-1} E^+_{-2} \, Q^{(-2)}_\ell \, E^-
E^- E^- E^- \, .
\end{equation}
Analogous statements can be made for $s=-1$ and $s=1$.

\newpage
\section{Supersymmetric Structure}
\label{SUSY}

The $\ell$-ladder structure, with the equation of motion operator
$H_\ell$ written as some
sort of square, is reminiscent of supersymmetric quantum mechanics.
The supersymmetric structure can be made explicit as follows.
It is convenient to define $A_\ell$ and $A_\ell^\dagger$:
\be
\begin{aligned}
 A_\ell &\equiv D^-_\ell = \Delta \partial_z + {1\over 2} \big[\ell(z_k -
   2z) + 2iq z_k\big] \,  \\
 A_\ell^\dagger &\equiv D^+_{\ell-1}= - \Delta \partial_z + {1\over 2} \big[\ell(z_k - 2z) -
   2iq z_k\big] \ \, .
\end{aligned}
\ee
where we have used $D^\pm$ defined in eq. (\ref{DKerr}). 
The operator $A_\ell^\dagger$ is the hermitian conjugate of $A_\ell$
in the sense that the action can be thought of as $\int \rd y \,
\psi_\ell^* H_\ell \psi_\ell$ with $\partial_{y} \equiv
\Delta \partial_z$. 
Noting that 
\be
\begin{aligned}
A_\ell^\dagger A_\ell &= H_\ell + \frac{(\ell^2 + 4q^2)z_k^2}{4}\quad
, \quad
A_\ell A_\ell^\dagger &= H_{\ell-1} + \frac{(\ell^2 + 4q^2)z_k^2}{4},
\end{aligned}
\ee
we are motivated to define the operator
\be
{\cal H}_\ell \equiv 
\left( \begin{array}{cc} A_\ell^\dagger A_\ell & 0 \\ 0 & A_\ell
 A_\ell^\dagger 
\end{array}
\right) \, ,
\ee
such that its eigenvectors $\Psi_\ell$ satisfying
\be
{\cal H}_\ell \Psi_\ell
= E_\ell \Psi_\ell \,, \qquad
  E_\ell \equiv \frac{(\ell^2 + 4q^2) z_k^2}{4}\,,
\ee
consist precisely of the $\psi_\ell$ and $\psi_{\ell-1}$ solutions
packaged together. In particular, it is simple to check that two possible
solutions are:
\be
\Psi_\ell = \left( \begin{array}{c} \psi_\ell \\ 0 \end{array} \right)
  \,, \qquad \Psi_\ell = \left( \begin{array}{c} 0 \\ A_\ell
                                     \psi_\ell \end{array} \right) \, ,
\ee
where $\psi_\ell$ solves $H_\ell \psi_\ell = 0$ and $A_\ell \psi_\ell$
solves $H_{\ell-1} (A_\ell \psi_\ell) = 0$. The operator ${\cal H}_\ell$ can be thought of as 
coming from an anti-commutator of ``supercharges"
\be
{\cal H}_\ell = \{ {\cal Q}_\ell , {\cal Q}_\ell^\dagger \} \,, \qquad {\rm with}\quad
{\cal Q}_\ell \equiv 
\left( \begin{array}{cc} 0 & 0 \\ A_\ell & 0 \end{array} \right) \, , 
\ee
which satisfy $[{\cal H}_\ell , {\cal Q}_\ell] = [{\cal H}_\ell, {\cal
  Q}_\ell^\dagger] =  0$. 
From this perspective, the Hamiltonian at level $\ell$ is the supersymmetric partner of the Hamiltonian at level $\ell+1$ and so on (indicating that these potentials are shape invariant). We can then understand the simplicity of the tower of potentials from the fact that the $\ell =0$ potential is just a constant.\footnote{This supersymmetric structure at $q=0$ is just the standard structure of the P\"oschl--Teller potential, discussed in Appendix~\ref{app:PT}.}
We have spelled out the supersymmetric structure
for the scalar in a Kerr background. The same can be done for the
Teukolsky equation, with conjugation involving $s \rightarrow -s$. 
Incidentally, in the context of metric perturbations around 
a Schwarzschild background, there is a different sort of
supersymmetry that has been discussed in the literature, one that
relates parity even and odd modes, 
obeying the Zerilli and 
Regge--Wheeler equations. The Teukolsky equation provides an
economical description, 
with the even and odd modes described by
the real and imaginary parts of the Newman--Penrose variable,
making the symmetry between them manifest.

\newpage
\section{Hypergeometric Identities}
\label{hyperg}

In uncovering the ladder structures in black hole perturbations, we
were aided by the known static solutions which are in the form of
the (Gauss) hypergeometric function $\hypergeom{2}{1}$.\footnote{For a review of some of the elementary properties of hypergeometric functions, see Appendix B of~\cite{Hui:2020xxx}.}
Neighboring hypergeometric functions, with
arguments offset by unity, are known to be related.
The identities we find the most helpful in revealing the $\ell$-ladder
are:
\begin{align}
\label{eq:hyper1}
\def\arraystretch{.8}
\begin{aligned}
\def\arraystretch{.8}
\frac{\rd}{\rd z} \hypergeom{2}{1}\left[\begin{array}{c}
a,~~b\\
c
\end{array}\Big\rvert \,z\,\right]  = ~&{a \big[ (c-a-1) - (b-a-1)(1-z) \big] \over
 z(b-a-1)(1-z)} \hypergeom{2}{1}\left[\begin{array}{c}
a,~~b\\
c
\end{array}\Big\rvert \,z\,\right] \\
&- {a (c-b) \over z (b-a-1)(1-z)} \hypergeom{2}{1}\left[\begin{array}{c}
a+1,~b-1\\
c
\end{array}\Big\rvert \,z\,\right]\,,
\end{aligned}\\[6pt]
\begin{aligned}
\def\arraystretch{.8}
\frac{\rd}{\rd z} \hypergeom{2}{1}\left[\begin{array}{c}
a,~~b\\
c
\end{array}\Big\rvert \,z\,\right]  = ~& {b \big[(c-b-1) - (a-b-1)(1-z)\big] \over z(a-b-1)(1-z)}  \hypergeom{2}{1}\def\arraystretch{.8}\left[\begin{array}{c}
a,~~b\\
c
\end{array}\Big\rvert \,z\,\right] \\
&- {b (c-a) \over z (a-b-1)(1-z)} \hypergeom{2}{1}\def\arraystretch{.8}\left[\begin{array}{c}
a-1,~b+1\\
c
\end{array}\Big\rvert \,z\,\right]\,.
\end{aligned}
\label{eq:hyper2}
\end{align}
%
Note that the independent solutions to the equation \eqref{tildeH} take the form of 
\be
\label{psi1psi2APP}
\psi_{1\, \ell }^{(s)} = \def\arraystretch{.8} \hypergeom{2}{1}\left[\begin{array}{c}
-\ell - s,~ \ell - s + 1\\
1 - s - 2iq
\end{array}\Big\rvert \,\frac{z}{z_k}\,\right]\,,
\quad
\psi_{2 \, \ell }^{(s)} = \Delta^s \E^{-2iq\ln[(z-z_k)/z]} \def\arraystretch{.8} \hypergeom{2}{1}\left[\begin{array}{c}
-\ell + s,~ \ell + s + 1\\
1 + s + 2iq
\end{array}\Big\rvert \,\frac{z}{z_k}\,\right]\,,
\ee
where $\psi_{1\, \ell }^{(s)}$ corresponds to the purely growing branch.
The identities~\eqref{eq:hyper1} and~\eqref{eq:hyper2} 
can be derived from known identities in the
literature:
\begin{align}
&
\begin{aligned}
\label{Id1a}
 (c-a-1) 
\def\arraystretch{.8}\hypergeom{2}{1}\left[\begin{array}{c}
a,~~b\\
c
\end{array}\Big\rvert \,z\,\right] = ~&(b-a-1) (1-z) \def\arraystretch{.8}\hypergeom{2}{1}\left[\begin{array}{c}
a+1,~b\\
c
\end{array}\Big\rvert \,z\,\right] \\
&+ (c-b) \def\arraystretch{.8}\hypergeom{2}{1}\left[\begin{array}{c}
a+1,~~b-1\\
c
\end{array}\Big\rvert \,z\,\right]\, ,\\
\end{aligned}\\[6pt]
&
\begin{aligned}
\label{Id1b}
\big[(a-b)(a-b-1)(1-z) + b(c-b-1)\big] 
\def\arraystretch{.8}\hypergeom{2}{1}\left[\begin{array}{c}
a,~~b\\
c
\end{array}\Big\rvert \,z\,\right] = ~&a(a-b-1)(1-z) \def\arraystretch{.8}\hypergeom{2}{1}\left[\begin{array}{c}
a+1,~b\\
c
\end{array}\Big\rvert \,z\,\right] \\
&+  b(c-a) \def\arraystretch{.8}\hypergeom{2}{1}\left[\begin{array}{c}
a-1,~~b+1\\
c
\end{array}\Big\rvert \,z\,\right]\, ,
\end{aligned}\\
&
\begin{aligned}
\label{Id2}
\def\arraystretch{.8}\hypergeom{2}{1}\left[\begin{array}{c}
a,~~b\\
c
\end{array}\Big\rvert \,z\,\right] = ~& \def\arraystretch{.8}\hypergeom{2}{1}\left[\begin{array}{c}
a+1,~b\\
c
\end{array}\Big\rvert \,z\,\right] -  \frac{bz}{c} \def\arraystretch{.8}\hypergeom{2}{1}\left[\begin{array}{c}
a+1,~~b+1\\
c+1
\end{array}\Big\rvert \,z\,\right]\, ,
\end{aligned}\\[6pt]
&
\begin{aligned}
\label{Id3}
\frac{\rd}{\rd z}\def\arraystretch{.8}\hypergeom{2}{1}\left[\begin{array}{c}
a,~~b\\
c
\end{array}\Big\rvert \,z\,\right] = \frac{ab}{c}\def\arraystretch{.8}\hypergeom{2}{1}\left[\begin{array}{c}
a+1,~b+1\\
c+1
\end{array}\Big\rvert \,z\,\right]\,.
\end{aligned}
\end{align}
Equations~\eqref{Id1a} and~\eqref{Id1b}, generalizing the Gauss contiguous
relations, can be found in 1.27 and 2.1 of \cite{ContigRakha}.
Equation~\eqref{Id2} can be found on p. 43 of \cite{Stalker}.
Equation~\eqref{Id3} is well known and can be found on p. 41 of \cite{Magnus}.
The identities useful for revealing the $s$-ladder are eq.~\eqref{Id3} and:
\be
 {\rd\over \rd z} \left(z^{c-1} (1-z)^{a+b-c} \def\arraystretch{.8}\hypergeom{2}{1}\left[\begin{array}{c}
a,~~b\\
c
\end{array}\Big\rvert \,z\,\right] \right) =   (c-1) z^{c-2} (1-z)^{a+b-c-1} \def\arraystretch{.8}\hypergeom{2}{1}\left[\begin{array}{c}
a-1,~b-1\\
c-1
\end{array}\Big\rvert \,z\,\right] ,
\ee
from p. 42 of \cite{Magnus}.

\section{Scalar-Gauss--Bonnet Theories}
\label{app:GB}

In this appendix, we consider the case of a scalar field coupled to the Gauss--Bonnet curvature invariant $R_{\rm GB}^2 \equiv R^{\mu\nu\rho\sigma}R_{\mu\nu\rho\sigma} - 4R^{\mu\nu}R_{\mu\nu} + R^2$. Such a coupling makes it possible to evade the no-hair theorem, sourcing a nonzero scalar profile around a black hole with a decaying  $1/r$ tail at large $r$~\cite{Sotiriou:2013qea}.
This example is an interesting laboratory that allows us to test the consequences of  our ladder symmetries for Love numbers and hair in an explicit microscopic  model.

Let us start by adding  a linear coupling  of the form ${\cal L}_{\text{GB}}\sim \alpha\phi R_{\rm GB}^2$ to the ultraviolet theory of a scalar field coupled to gravity. For simplicity, we shall treat the coupling $\alpha$ in ${\cal L}_{\text{GB}}$  perturbatively and look for a solution to the field equations  in powers of $\alpha$. At linear order in $\alpha$, we can ignore the backreaction of the metric (corrections to the metric enter  at  order $\alpha^2$) and study  the scalar's dynamics on a fixed Schwarzschild geometry.
The scalar's equation of motion is
\begin{equation}
\square \phi = - \alpha R_{\rm GB}^2 \, .
\label{eqGB}
\end{equation}
After decomposing $\phi$ in spherical harmonics,  the Gauss--Bonnet term in~\eqref{eqGB} plays the role of a  source for the $\ell=0$ mode, while the equations for the higher multipoles are unchanged at this order in $\alpha$ with respect to~\eqref{phieom}. In the static limit, the solution for the monopole that  is regular  at the horizon is 
\begin{equation}
\phi_{\ell=0}(r)  = \phi_{\infty} + \alpha\left( \frac{4}{r r_s}  + \frac{2}{r^2}  +\frac{4r_s}{3r^3} \right) + \mathcal{O}(\alpha^2) \, ,
\label{solphiGB}
\end{equation}
where $\phi_{\infty}$ is a constant, corresponding to the value of
$\phi(r)$ at $r=\infty$. The term $4\alpha/(rr_s)$ is the well-known
fall-off that corresponds to the scalar hair generated by the
Gauss--Bonnet coupling~\cite{Sotiriou:2013qea}. On the other hand, it
is clear from~\eqref{solphiGB} that there is no static response
induced by a constant tidal field, which would be proportional to
$\phi_{\infty}$. 

The above calculations can be understood in terms of 
 (exact and broken) symmetries in the following way. The point is that the
coupling to the Gauss--Bonnet operator is responsible for  an explicit
breaking of the $Q_0$ symmetry, so that hair is no longer forbidden.
At the level of the point-particle
EFT~\eqref{eq:pointscalaraction}, the breaking of $Q_0$ implies that
the  tadpole $g\phi$ coupling can appear
in the infrared
description and should therefore be included in the EFT. In particular---combined with the spherical symmetry of the system---it implies that if we  expand  $g\phi$ in spherical harmonics only the monopole is allowed by symmetries. And indeed this is precisely what the microscopic calculation shows: the Gauss--Bonnet coupling allows for nonzero monopole hair (but not for higher-multipole-type of hair).
Even though the coupling to the Gauss--Bonnet operator breaks $Q_0$ explicitly, one finds that  $Q_{\ell\geq1}$ are still symmetries of the theory, up to linear order in $\alpha$. This is enough to ensure the vanishing of the Love numbers with  $\ell\geq1$.\footnote{Note that this result follows from knowing the exact microscopic dynamics. Formulating a similar statement directly at the level of the EFT~\eqref{eq:pointscalaraction} seems  more subtle. 
 Indeed, the breaking of  $Q_0$ allows in general  all the
 $\lambda_\ell$'s in~\eqref{eq:pointscalaraction}. Presumably,
 understanding how to implement a subset of our symmetries (at
 selected $\ell$'s) at the level of the worldline action~\eqref{eq:pointscalaraction}, as mentioned in Section~\ref{IRsymmetries}, would allow us to use the remaining   symmetries to constrain (at least some of) the other couplings $\lambda_\ell$. Nevertheless, the vanishing of the monopole Love number can at least be   understood in terms of   the shift symmetry $\phi\mapsto\phi+c$, which clearly forbids $\phi^2$ in~\eqref{eq:pointscalaraction}, providing in the infrared a symmetry explanation of why $\lambda_0=0$.}
On the other hand, the absence of a static response for $\ell=0$ is enforced by the shift symmetry, $\phi\mapsto\phi+c$, which is an exact invariance of the theory with linear coupling ${\cal L}_{\text{GB}}\sim \alpha\phi R_{\rm GB}^2$, {if we ignore any  backreaction on the metric (decoupling limit)}. As a result, all the black hole's scalar  Love numbers vanish. 

It is instructive to contrast  this result  with the case of a scalar
field coupled quadratically   to the Gauss--Bonnet operator, i.e.,
${\cal L}_{\text{GB}}\sim \frac{1}{2}\alpha\phi^2 R_{\rm GB}^2$
\cite{Silva:2017uqg}. In contrast to the previous example, the shift symmetry is now broken explicitly in the ultraviolet theory. Therefore, there is no reason now for the monopole Love number to vanish and  the quadratic worldline operator $\phi^2$  not to be allowed  in the point-particle effective description. This is indeed what one finds from an explicit calculation in the microscopic theory and after matching with the EFT. To show this, we again solve the scalar equation of motion, which now reads
\begin{equation}
\square \phi + \alpha \phi R_{\rm GB}^2 =0  \, ,
\label{eqGB-2}
\end{equation}
perturbatively in $\alpha$. Focusing for simplicity on the monopole $\ell=0$, the solution to~\eqref{eqGB-2} that is regular at $r=r_s$ is now
\begin{equation}
\phi_{\ell=0}(r)  = \phi_{\infty} \left( 1 + \alpha \left( \frac{4}{r r_s}  + \frac{2}{r^2}  +\frac{4r_s}{3r^3} \right) + \mathcal{O}(\alpha^2) \right)  \, ,
\label{solphiGB-2}
\end{equation}
where we have discarded terms that are order $\alpha^2$ or higher. The
constant $\phi_{\infty}$ now multiplies the whole solution---this
follows from the fact that, as opposed to~\eqref{eqGB},
eq.~\eqref{eqGB-2} is homogeneous in $\phi$---and one can genuinely
interpret the $1/r$ tail as the static response to the monopole tidal
field profile $\phi_{\infty}$. According to 
\cite{Silva:2017uqg}, this theory can even generate hair,
which from the EFT perspective is described by a linear $\phi$
coupling to the worldline. This represents a spontaneous breaking
of the discrete $\phi \rightarrow -\phi$ symmetry that is present in
the microscopic theory. This so called scalarization phenomenon
seems to occur only if $\alpha$ is sufficiently large.

\vspace{1.0cm}

\ornamentSep

\vspace{1.0cm}

Our exploration of the black hole ladders---black hole arpeggios as it were---was partly inspired by Bach's music. A  ladder structure can be
discerned for instance in the celebrated
prelude of Cello Suite No.~1 in G major.
\begin{equation*}
\includegraphics[width=0.5\textwidth]{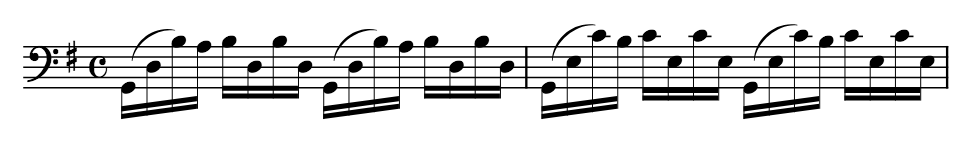} 
\end{equation*}

\vspace{-0.5cm}

\newpage
\phantomsection
\addcontentsline{toc}{section}{References}
\bibliographystyle{utphys}
{\small
\bibliography{notesbib}
}

\end{document}